\documentclass[prfluids,preprint]{revtex4-2}

\usepackage{graphicx}
\usepackage{dcolumn}
\usepackage{bm}
\usepackage{xcolor}
\usepackage[colorlinks=true]{hyperref}
\usepackage{amsmath}
\usepackage{graphicx}
\usepackage{bm}
\usepackage{dcolumn}

\begin{document}


\title{
Full Turbulence Simulation of Channel Flow at $Re_{\tau} \approx 1000$
}

\author{Yoshinobu Yamamoto}
 \email{yamamotoy@yamanashi.ac.jp}
 \affiliation{%
 Department of Mechanical Engineering, University of Yamanashi, 4-3-11, Takeda, Kofu 400-8511, Japan
}%
\author{Yoshiyuki Tsuji}%
 \email{c42406a@cc.nagoya-u.ac.jp}
\affiliation{%
Department of Energy Science and Engineering, Nagoya University, Furo, Chikusa, Nagoya 464-8601, Japan
}%

\date{\today}

\begin{abstract}
A Full Turbulence Simulation (FTS) of turbulent channel flow at friction Reynolds number ($Re_{\tau}$) $\approx 1000$ was performed by resolving the Kolmogorov wavenumber in all spatial directions. At this Reynolds number, the intermediate layer attains a physically meaningful width and is fully resolved in the present computation, providing the reference dataset that captures its turbulence and dissipation characteristics with high fidelity. The wall‑normal grid spacing of the FTS also confirms that, when the Kolmogorov length scale is sufficiently resolved, the second‑order central‑difference scheme introduces no adverse numerical effects in the wall‑normal direction. In the wall‑parallel directions, two resolution criteria were identified based on the present FTS: a first‑approximation DNS resolution that resolves more than 99\% of the turbulent kinetic energy and dissipation rate ($\Delta x^{+} \approx 19$, $\Delta y^{+} \approx 8$, where $\Delta x^{+}$ and $\Delta y^{+}$ denote the stream and spanwise spatial resolutions in wall units) and a full dissipation‑resolution criterion ($\Delta x^{+} \approx 7.5$, $\Delta y^{+} \approx 5.0$). The first‑approximation resolution by means of spectral method reproduces the essential turbulence statistics within 1\% accuracy while requiring only one‑eighth of the grid points used in the FTS, demonstrating its practical efficiency. In contrast, even the highest‑resolution second‑order central‑difference case ($\Delta x^{+} \approx 5.0$, $\Delta y^{+} \approx 4.5$) fails to match the accuracy of the first‑approximation spectral resolution.
These findings provide important resolution guidelines for high‑Reynolds‑number DNS, particularly for simulations at $Re_{\tau} = \mathcal{O}(10^{4})$ 
\end{abstract}

\maketitle

\section{\label{sec:level1}INTRODUCTION}
Since the first Direct Numerical Simulation (DNS) of Homogeneous Isotropic Turbulence (HIT)\cite{orszag1972numerical}, DNS has been established as one of the most powerful tools for verifying turbulence theory, modeling, and improving the accuracy of experimental techniques. In general, DNS requires spatial grid resolution capable of resolving the smallest eddies down to the Kolmogorov length scale ($\eta \equiv (\nu^{3}/\varepsilon)^{1/4}$, where  $\varepsilon$ is the turbulent energy dissipation rate per unit mass, $\nu$ is the kinematic viscosity). In actual DNS of HIT, the grid resolution corresponding to Kolmogorov wavenumbers ($k_{max}\eta =1$ where $k_{max}$ is the maximum wave number) or higher ($k_{max}\eta =4$) is applied (see e.g., \cite{yokokawa200216, ishihara2016energy, yeung2018effects, yeung2025gpu}). In contrast, Turbulent Channel Flow (TCF), which is anisotropic, commonly uses spatial resolutions that are coarser than the Kolmogorov scale and even below the Kolmogorov wavenumber. This practice has been considered reasonable based on previous DNS results(e.g., \cite{kim1987turbulence, tanahashi2004scaling}), which show that high-wavenumber attenuation in the dissipation energy spectrum is sufficiently achieved at Kolmogorov wavenumber. However, as the friction Reynolds number ($Re_{\tau} \equiv h/l_{\tau}$ where $h$ is channel half height, $l_{\tau} \equiv \sqrt{\tau_{w}/\rho}$ is the viscous length scale,velocity, $u_{\tau} \equiv \sqrt{\tau_{w}/\rho}$ is the friction velocity, $\tau_{w}$ is the wall shear stress, and $\rho$ is the density) increases, it is not necessarily obvious whether a spatial resolution equivalent to that at low Reynolds numbers can be applied. In fact, as $Re_{\tau}$ increases, the existence of the intermediate layer exhibiting transitional scaling has been observed, in addition to the outer layer that scales with $h$ and the near-wall region that scales with $l_{\tau}$(see e.g., \cite{afzal1982fully, klewicki2007physical, mckeon2017engine}). This intermediate layer corresponds to the wall-normal height where the universal scaling law of turbulence statistics is expected to hold (see, e.g., \cite{hultmark2012turbulent, marusic2013logarithmic}), and represents this region requiring accurate reproducibility in high-Reynolds-number DNS.
Recent studies suggest the presence of an intermediate layer exhibiting transitional scaling between the inner and outer regions. Although its precise definition is still under active discussion, several works \cite{klewicki2010reynolds, klewicki2021properties} indicate that this region may emerge approximately within
\begin{equation}
1.5Re_{\tau}^{1/2} \le z^{+} \le 0.15\,Re_{\tau},
\label{eq:range}
\end{equation}
where $z^{+}$ denotes the wall-normal distance normalized by $l_{\tau}$. At relatively low Reynolds numbers, this range becomes extremely narrow. For example, $Re_{\tau} =180: 20 < z^{+} < 27, Re_{\tau}=395: 30 < z^{+} < 60$ providing only a limited wall-normal extent in which transitional scaling can be observed. A sufficiently large Reynolds number is therefore required to ensure that the intermediate layer is wide enough to be meaningfully resolved. For instance, securing more than 100 viscous units within this region already requires $Re_{\tau} \ge 1000$. This consideration motivates performing FTS at $Re_{\tau} \approx 1000$, which represents the minimum Reynolds number at which the intermediate layer becomes sufficiently developed for quantitative assessment. As a reference for DNS that resolve the Kolmogorov wavenumber in all directions at $Re_{\tau} \ge 1000$, we note the thermal channel–flow DNS by Alc\'antara-\'Avila \& Hoyas~\cite{alcantara2021direct}. Their resolution strategy is informative for the present study. It should be noted, however, that their simulations employed a relatively small computational domain, and some caution is therefore appropriate when interpreting their results, particularly regarding outer-layer statistics.
.

In addition to the physical motivation described above, there is a practical computational motivation for establishing reliable resolution criteria. Modern supercomputers achieve extremely high floating‑point throughput but often have limited bisection bandwidth, making communication a dominant bottleneck in large‑scale DNS. Consequently, performing DNS at  inevitably requires certain compromises—such as the use of moderately resolved grids or the adoption of second‑order central‑difference (CD2) schemes—to maintain computational efficiency. Recent studies have raised concerns regarding the applicability of CD2 schemes at high Reynolds numbers (e.g., \cite{yao2023direct,nagib2024utilizing}). However, the suitability of a discretization method must be assessed relative to the grid resolution at which it is applied.
It is not appropriate to assume that spectral methods are always more accurate or that CD2 schemes are inherently unreliable; a CD2 scheme operating on a grid finer than the effective resolution of a spectral method can, in principle, yield higher accuracy. Thus, what is needed is a resolution‑based evaluation of numerical accuracy. A second objective of the present study is therefore to use the fully resolved FTS at  as a reference for quantitatively assessing the accuracy of CD2 schemes and moderately resolved DNS.
This provides a practical foundation for designing high‑Reynolds‑number DNS on modern supercomputing architectures.

Since performing FTS at the world's highest Reynolds numbers is currently impractical due to computational resource limitations, it is essential to conduct FTS at conditions where it is still feasible but sufficiently high to capture the transitional scaling. In this study, Full Turbulence Simulation (FTS) resolving the Kolmogorov wavenumber in all directions is conducted for channel flow at $Re_{\tau} \approx 1000$. Based on the FTS results, the required spatial resolution—defined as the grid spacing that reproduces the mean velocity, turbulent kinetic energy, and dissipation rate within 1\% error—is quantitatively evaluated, thereby providing a foundation for validating DNS at$Re_{\tau} = \mathcal{O}(10^{4})$. 

The structure of this paper is as follows. Chapter \ref{sec:level2} describes the numerical method and validates the present code. Chapter \ref{sec:level3} examines the computational domain and spatial resolution required for FTS. Chapter \ref{sec:level4} presents the FTS at $Re_{\tau} \approx 1000$ and analyzes its turbulence statistics. Chapter \ref{sec:level5} evaluates the resolution criteria derived from the FTS. Chapter \ref{sec:level6} assesses the accuracy of the second‑order central difference scheme. Chapter \ref{sec:level7} summarizes the findings and discusses implications for high‑Reynolds‑number DNS.

\section{\label{sec:level2}Overview of The Current DNS Code}
\subsection{Numerical formulation}

\begin{figure*}
\centering
\includegraphics[width=.7\textwidth]{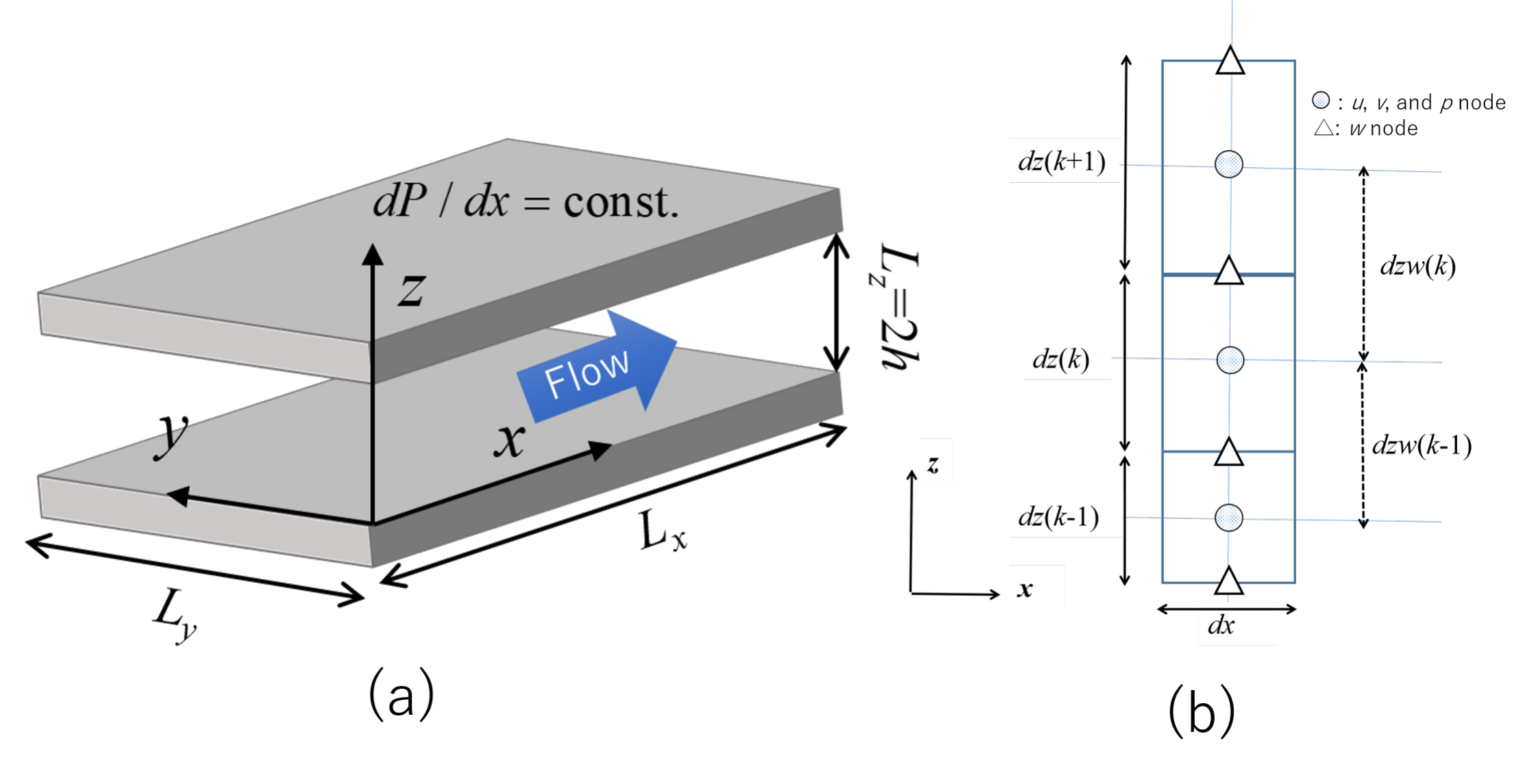}
\caption[ ]{
(a)Simulation geometry and (b) wall-normal staggered grid system in turbulent channel flow (TCF).
}
\label{fig:TCF-Z}
\end{figure*}

The simulated case in TCF, which is the flow between parallel planes driven by the pressure gradient ($f$) to maintain the constant mass flow rate, as shown in Fig.\ref{fig:TCF-Z} (a). The boundary conditions include periodic boundary conditions in the streamwise ($x$) and spanwise ($y$) directions, and non-slip (zero velocity) boundary condition at the walls ($z =0$ and $2h$). The system of the unsteady Navier–Stokes equations (\ref{eq:NS}) for incompressible fluids in three-dimensions and skew-symmetric form and continuity equation (\ref{eq:Mass}), is considered ($i, j = 1, 2, 3$):
\begin{equation}
\frac{\partial u_i}{\partial t}
=
\underbrace{-
\frac{1}{2}
\left(
    \frac{\partial (u_i u_j)}{\partial x_j}
    +
    u_j \frac{\partial u_i}{\partial x_j}
\right)
+
f\,\delta_{i1}
+
\nu \frac{\partial^2 u_i}{\partial x_j \partial x_j}
}_{H_i}
-
\frac{\partial}{\partial x_i}\left(\frac{p}{\rho}\right),
\label{eq:NS}
\end{equation}
\begin{equation}
\frac{\partial u_i}{\partial x_i}= 0.
\label{eq:Mass}
\end{equation}

Here $u_{i} = (u, v, w)$ are the velocity components in the Cartesian coordinates $x_{i} = (x, y, z)$, and $p$ is the pressure, respectively. The time integration method used to solve Eqs. (\ref{eq:NS}) and (\ref{eq:Mass}) is based on the fractional step method employing the Euler implicit scheme for pressure term and the second-order-explicit Adams-Bashforth scheme for the other terms ($H_i$). Thus, Eqs. (\ref{eq:NS}) and (\ref{eq:Mass}) can be written as
\begin{equation}
\frac{u_i^{F} - u_i^{n}}{\Delta t}
= \frac{3}{2} H_i^{n} - \frac{1}{2} H_i^{n-1},
\label{eq;NS-AB}
\end{equation}
\begin{equation}
\frac{u_i^{n+1} - u_i^{F}}{\Delta t}
= -\frac{\partial}{\partial x_i}
\left( \frac{p^{n+1}}{\rho} \right),
\label{eq:NS-FS}
\end{equation}
with
\begin{equation}
\frac{\partial u_i^{n+1}}{\partial x_i} = 0.
\label{eq:Mass-FS}
\end{equation}
Here $\Delta t$ is the time-step width, and super-script $n$ and $F$ denotes time-step level. Taking the divergence of Eqs (\ref{eq:NS-FS}) and considering Eq. (\ref{eq:Mass-FS}), we obtain the Poisson equation (\ref{eq:PPE}) for the pressure at the new time-step level: $n +1$ as follow:
\begin{equation}
\frac{\partial^2}{\partial x_i \partial x_i}
\left( \frac{p^{n+1}}{\rho} \right)
= \frac{1}{\Delta t}
\frac{\partial u_i^{F}}{\partial x_i}
\equiv D.
\label{eq:PPE}
\end{equation}
For spatial discretization, the Fourier spectral method is applied in the $x$ and $y$ directions. Accordingly, the fields in these directions in Eqs. (\ref{eq:NS}) and (\ref{eq:Mass}) are Fourier transformed. The nonlinear terms in Eqs. (\ref{eq:NS}) are evaluated pseudo-spectrally by transforming the velocities back to physical space to compute their product, using Fast Fourier Transform (FFT). Although the use of the pseudo-spectral method introduces aliasing errors, it is known that these errors can be reduced by applying the skew-symmetric form to the nonlinear terms (e.g. \cite{kravchenko1997effect}). Therefore, in this code, aliasing errors are not removed; instead, the high-wavenumber cutoff filter in Eq. (\ref{eq:filter}) is multiplied with the nonlinear terms (see e.g., \cite{hou2007computing}).

\begin{equation}
\begin{aligned}
r_x(k_x) &=
\left\{
\begin{array}{ll}
1, & \text{if } \left| \frac{2 k_x L_x}{2\pi N_x} \right| \le C_a, \\[6pt]
0, & \text{if } \left| \frac{2 k_x L_x}{2\pi N_x} \right| > C_a,
\end{array}
\right.
\\[10pt]
r_y(k_y) &=
\left\{
\begin{array}{ll}
1, & \text{if } \left| \frac{2 k_y L_y}{2\pi N_y} \right| \le C_a, \\[6pt]
0, & \text{if } \left| \frac{2 k_y L_y}{2\pi N_y} \right| > C_a,
\end{array}
\right.
\end{aligned}
\label{eq:filter}
\end{equation}

where, $k_x$ and $k_y$ represent the wavenumbers, while $L_{x} (N_{x})$ and $L_{y} (N_{y})$ denote the computational domain sizes (grid numbers in real space) in the $x$ and $y$ directions, respectively, $C_a$ represents the threshold, where applying $C_{a} = 2/3$ corresponds to implementing the two-thirds rule, which completely removes aliasing errors. In this study, the value of 62/64 is applied, meaning that when using the 64-point grid, the effective number of grid points is 62. The effects of this approach will be examined in Chapter \ref{sec:level5}.
Meanwhile, the second-order central difference method is employed for the wall-normal direction. The staggered grid system was employed, in which the velocity component in the only wall-normal direction was shifted by half a grid as shown in Fig.\ref{fig:TCF-Z}(b). The grid configuration in the wall-normal direction was based on a hyperbolic tangent function, resulting in a fine grid near the wall $(z =0, 2h)$ and a coarse grid at the center of the channel $(z =h)$. When the Fourier spectral method is applied in the $x$ and $y$ directions and the second-order accurate central difference method is applied in the $z$ direction, the discretized form of Eq. (\ref{eq:PPE}) is represented as a tri-diagonal matrix in the $z$ direction for pressure, as written in Eq. (\ref{eq:TDMA}):

\begin{equation}
A(l,m,k)\,\hat{p}(l,m,k) = B(l,m,k)\,\hat{p}(l,m,k+1) + C(l,m,k)\,\hat{p}(l,m,k-1) + \hat{D}(l,m,k),
\label{eq:TDMA}
\end{equation}
here
\begin{equation}
A(l,m,k)=\frac{1}{dz(k)}\left[  \frac{-1}{dzw(k)} +  \frac{-1}{dzw(k-1)} \right] - \left( k_{x}^2 + k_{y}^2 \right),
\label{eq:A}
\end{equation}
\begin{equation}
B(l,m,k)=\frac{1}{dz(k)dzw(k)},
\label{eq:B}
\end{equation}
\begin{equation}
C(l,m,k) = \frac{1}{dz(k)dzw(k-1)}.
\label{eq:C}
\end{equation}
In these expressions, the hat symbol denotes a Fourier transform value, while $dz$ and $dzw$ correspond to the grid widths in the $z$ direction as shown in FIG.\ref{fig:TCF-Z}(b), whereas $l$, $m$, and $k$ denote the grid number indices in the $k_x$, $k_y$, and $z$ directions, respectively. The tri-diagonal matrix in Eq. (\ref{eq:TDMA}) can be solved efficiently using the Tri-Diagonal Matrix Algorithm (TDMA)\cite{thomas1949elliptic}.
The time-step width $\Delta t$ is set such that $C_t$ in Eq. (\ref{eq:CFL}) never exceeds 0.25.
\begin{equation}
C_t = \max\left(
\Delta t \left| \frac{u}{\Delta x} \right|,
\; \Delta t \left| \frac{v}{\Delta y} \right|,
\; \Delta t \left| \frac{w}{\Delta z} \right|
\right).
\label{eq:CFL}
\end{equation}
The current code can also be modified to use the second- or third-order Runge-Kutta time integration scheme. However, under the aforementioned conditions, no significant differences were observed in the turbulence statistics. Even when using the Adams-Bashforth scheme, the second-order Runge-Kutta scheme is applied in the first step.

\subsection{Validation of the current code}

\begin{table}[t]
\centering
\scriptsize
\caption{Validation of the current code in comparison with DNS\cite{lee2015direct}.}

\begin{tabular}{lccccccccccc}
\hline
RUN & discretization ($x,y$) & discretization ($z$) & aliasing error &
$Re_{\tau}$ & $Re_b$ & $L_x/h$ & $L_y/h$ &
$N_x$ ($\Delta x^+$) & $N_y$ ($\Delta y^+$) & $N_z$ ($\Delta z^+$) & $T^+/Re_{\tau}$ \\
\hline
AH\cite{alcantara2021direct} &
Fourier & compact FD & 3/2 rule &
1019 & 22300 & $2\pi$ & $\pi$ &
NA (4.1) & NA (2.1) & 632 (0.16--4.5) & 86 \\
LM\cite{lee2015direct} &
Fourier & b-spline & 3/2 rule &
1000 & 20000 & $8\pi$ & $3\pi$ &
2304 (10.9) & 2048 (4.6) & 512 (0.02--6.2) & 12.5 \\
R1000L &
Fourier & CD2 & aliased &
1000 & 20000 & $8\pi$ & $3\pi$ &
2304 (10.9) & 2048 (4.6) & 512 (0.6--8.0) & 25.0 \\
\hline
\end{tabular}

\vspace{4pt}
\begin{flushleft}
\scriptsize
Here, $Re_{\tau}=u_{\tau}h/\nu$ is the friction Reynolds number,
$u_{\tau}$ the friction velocity, $h$ the channel half width, and $\nu$ the kinematic viscosity.
$Re_b = U_b h/\nu$ is the bulk Reynolds number, where $U_b$ is the bulk mean velocity.
$L_x$ and $L_y$ denote the streamwise and spanwise computational lengths.
$N_x$ ($\Delta x^+$), $N_y$ ($\Delta y^+$), and $N_z$ ($\Delta z^+$) are the grid numbers
in the streamwise, spanwise, and wall-normal directions, respectively.
The superscript $+$ denotes normalization using the friction length $l_{\tau}=\nu/u_{\tau}$
and friction velocity $u_{\tau}$.
\end{flushleft}
\label{tab:LM}
\end{table}

\begin{figure}
\centering
\includegraphics[width=.6\textwidth]{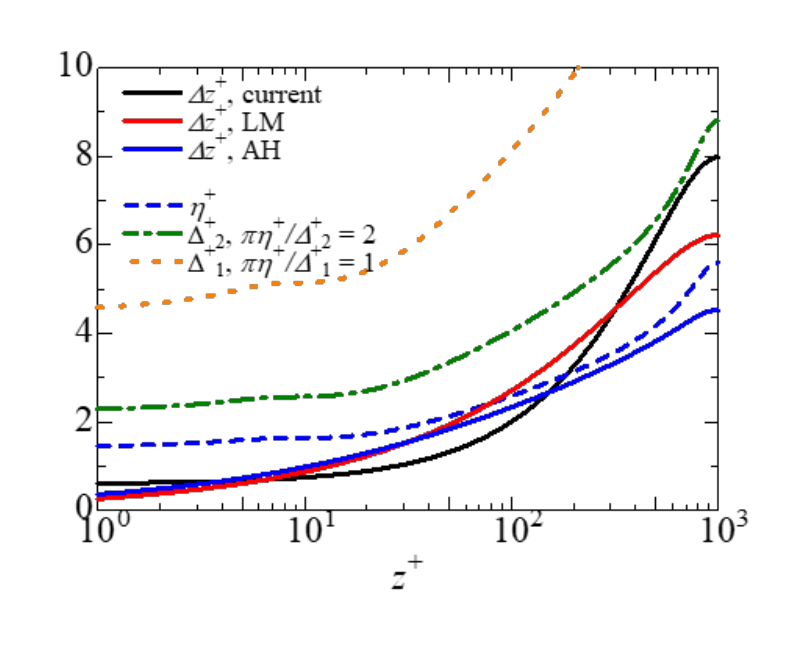}
\caption[ ]{
wall-normal grid resolution in R1000L.
}
\label{fig:kizamiz}
\end{figure}

In this chapter and thereafter, the averages are taken over the homogeneous directions (i.e. the streamwise- and spanwise-directions) and time. The corresponding values are displayed with the superscript bar ($-$). The mean flow $\boldsymbol{U}$ is unidirectional and corresponds to the form $\boldsymbol{U} = (U(z), 0, 0)$ in the Cartesian coordinate system $x = (x_{1}, x_{2}, x_{3}) = (x, y, z)$, and the fluctuating field $\boldsymbol{u}= (u_{1}, u_{2}, u_{3}) = (u, v, w)$ is statistically homogeneous in the $x$(streamwise)- and $y$(spanwise)-directions. Note that the velocity $\boldsymbol{u}$ in the Navier-Stokes Eq.(\ref{eq:NS}) is replaced with above $\boldsymbol{U} + \boldsymbol{u}$ as described above. Under the assumptions of statistical stationarity and homogeneity in the $x$- and $y$-directions of the TCF, the average of turbulence statistics is independent of time and depends on the position vector $\boldsymbol{x}$ only through $z$. The superscript $+$ denotes non-dimensionalization using the friction length $l_{\tau} (=\nu/u_{\tau})$ and friction velocity $u_{\tau}$.
The current code is applied under the same computational conditions as the previous DNS study \cite{lee2015direct} to demonstrate its validity. Table \ref{tab:LM} summarizes the computational parameters, including edomain size and grid resolution, which are matched to those in \cite{lee2015direct}.
Table \ref{tab:LM} also lists the computational parameters reported in Alc\'antara-\'Avila \& Hoyas~\cite{alcantara2021direct}. Their simulation was performed at a bulk Reynolds number about 10\% higher than that of LM\cite{lee2015direct}
($Re_{b}=20000$), and employed a smaller computational domain. In addition, the reported wall-parallel grid spacings do not clarify whether they correspond to the
effective wavenumber resolution or on the collocation grid associated with the dealiased velocity field (3/2 rule). For these reasons, their parameters are included only for reference. The primary difference lies in the wall-normal discretization, where the grid arrangement differs slightly, as shown in Fig.\ref{fig:kizamiz}.
In Fig.\ref{fig:kizamiz}, $\Delta_{1}$ and $\Delta_{2}$ denote the grid resolutions required to resolve the Kolmogorov wavenumber and twice that value, respectively, defined by
\begin{equation}
\frac{2\pi}{2\Delta_1}\!=\!\frac{1}{\eta},\;
\frac{2\pi}{2\Delta_2}\!=\!\frac{2}{\eta},\;
\eta\!=\!\left(\frac{\nu^{3}}{\varepsilon}\right)^{1/4},\;
\varepsilon\!=\!\nu\,\overline{\left(\frac{\partial u_i}{\partial x_j}\right)^2}.
\label{eq:eta}
\end{equation}
\begin{figure}
\centering
\includegraphics[width=.9\textwidth]{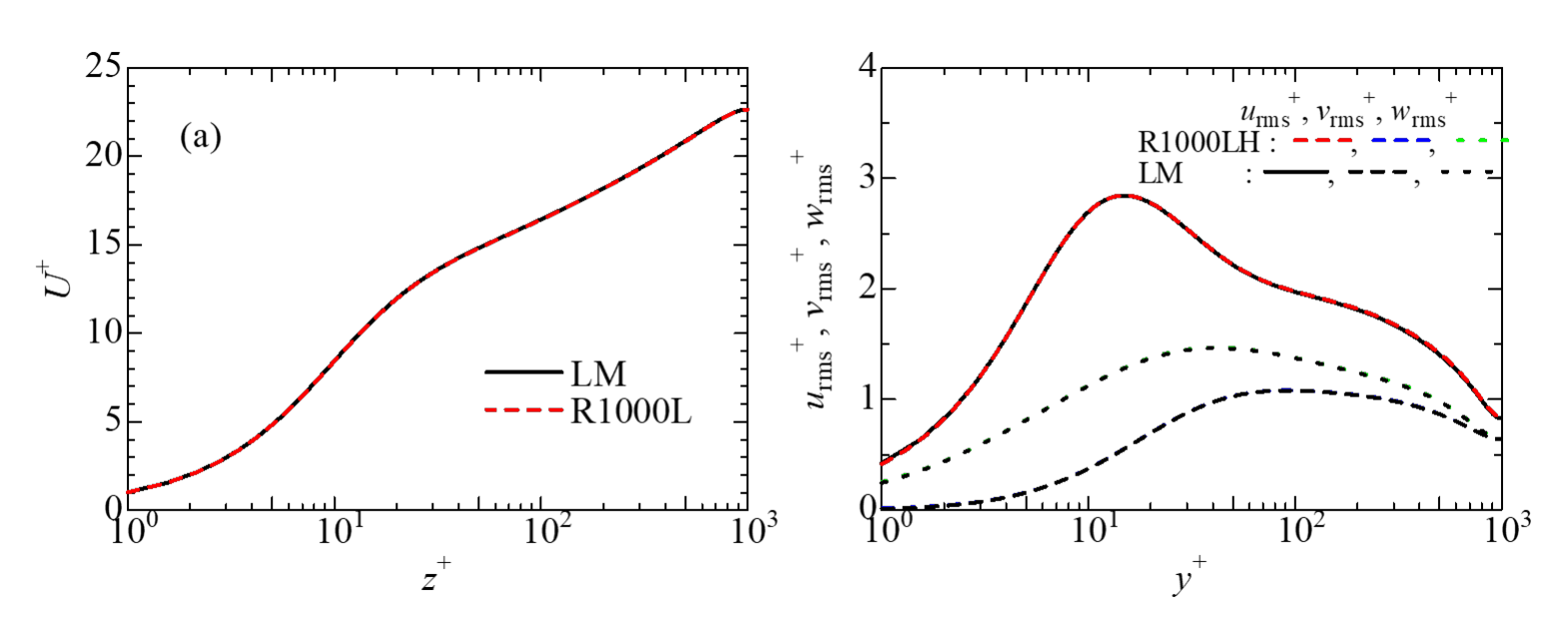}
\caption[ ]{
Comparison between R1000L and LM data\cite{lee2015direct} for the streamwise mean velocity and turbulent intensities: (a) Streamwise mean velocity, (b) Turbulent intensities.
}
\label{fig:U-rms-LM}
\end{figure}
\begin{figure}
\centering
\includegraphics[width=.9\textwidth]{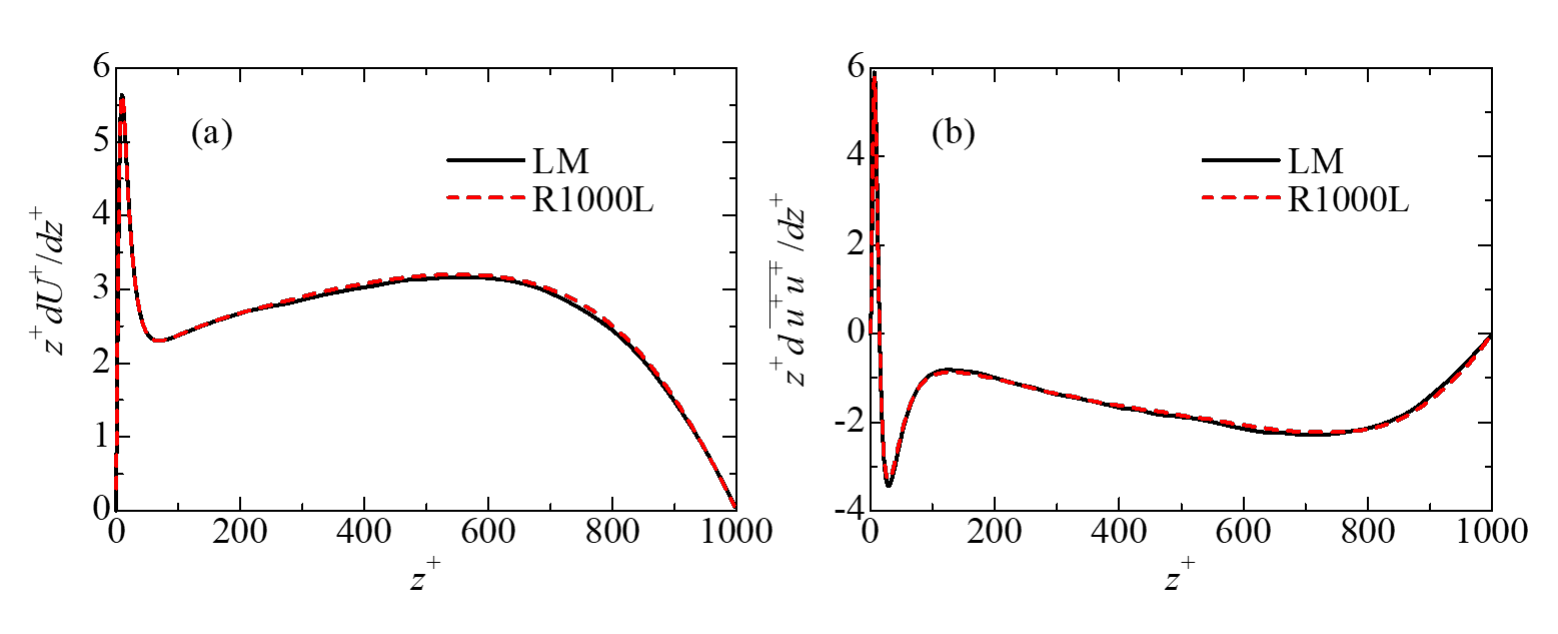}
\caption[ ]{
Comparison between R1000L and LM data\cite{lee2015direct} for pre-multiplied profiles of wall-normal derivative of streamwise mean velocity and turbulent intensities: (a) Streamwise mean velocity, (b) Streamwise normal Reynolds stress.
}
\label{fig:z-dUdz-duudz-LM}
\end{figure}
As shown in Fig.\ref{fig:kizamiz}, the wall-normal grid spacing $\Delta z$ in the R1000L case is sufficiently fine to resolve the Kolmogorov length scale ($\eta$). In particular, over the range $z^{+} < 300$ in Eq. (\ref{eq:range}) with $Re_{\tau} \approx 1000$, which fully covers the intermediate layer, the present grid satisfies $\Delta z \ll \eta$.The first off-wall grid point is located at $z_{1}^{+} \approx 0.3$ since $\Delta z^{+}\approx 0.6$ as shown in Fig.\ref{fig:kizamiz}(b), satisfying the stricter recommendation $z_{1}^{+} < 0.5$ proposed by Pirozzoli \& Orlandi\cite{pirozzoli2021natural}.
Furthermore, the $\Delta_2$ criterion (green curve), corresponding to $\Delta_{2} \approx \pi \eta/2$, represents the sufficient resolution level suggested by Pirozzoli \& Orlandi\cite{pirozzoli2021natural}. The present $\Delta z^{+}$ distribution (black curve) lies well below this threshold across the entire wall-normal range, indicating that the wall-normal resolution is significantly finer than the recommended requirement. Therefore, the wall-normal discretization in the R1000L case is sufficiently fine and does not limit the accuracy of the present validation.
The deviation of the total shear stress from its ideal linear distribution is extremely small across all simulations. As shown in Appendix \ref{app:convergence}, the maximum departure from linearity is on the order of $10^{-3}$ in wall units, indicating that the mean momentum balance is fully converged and that the statistical sampling is sufficient for all cases considered.
Figure \ref{fig:U-rms-LM} shows the profiles of streamwise mean velocity and turbulent intensities in comparison with LM\cite{lee2015direct}. As a more detailed comparison, Figure \ref{fig:z-dUdz-duudz-LM} presents the wall-normal derivative of the streamwise mean velocity and normal stress, both multiplied by the wall-normal height $z$. These statistics play an important role in validating turbulence theories in wall-turbulence. The results of the current code are confirmed to be in complete agreement with those of LM\cite{lee2015direct}, as shown in Figs. \ref{fig:U-rms-LM} and \ref{fig:z-dUdz-duudz-LM}.
It should be emphasized that the present numerical scheme—consisting of an explicit time integration method, second-order central difference in the wall-normal direction, and pseudo-spectral treatment without full aliasing removal—is fully validated under the condition that the grid resolution resolves the Kolmogorov length scale and the CFL number remains below 0.25. As demonstrated in the comparisons with Lee \& Moser \cite{lee2015direct}, these choices do not compromise the accuracy of turbulence statistics, and thus require no further justification within the scope of this study.

\section{\label{sec:level3}Estimation of Spatial Resolution and Domain Size for FTS}
\begin{table}[t]
\centering
\scriptsize
\caption{Computational domain and grid-resolution settings for the reduced-domain DNS (R1000) and the fully resolved reference simulation (FTS).}

\begin{tabular}{lccccccccccc}
\hline
RUN & discretization ($x,y$) & discretization ($z$) & aliasing error &
$Re_{\tau}$ & $Re_b$ & $L_x/h$ & $L_y/h$ &
$N_x$ ($\Delta x^+$) & $N_y$ ($\Delta y^+$) & $N_z$ ($\Delta z^+$) & $T^+/Re_{\tau}$ \\
\hline
R1000 &
Fourier & CD2 & aliased &
1000 & 20000 & 16 & 6.4 &
1600 (10.0) & 1296 (4.9) & 512 (0.6--8.0) & 25.0 \\
FTS &
Fourier & CD2 & aliased &
999 & 20000 & 16 & 6.4 &
4000 (4.0) & 1600 (4.0) & 512 (0.6--8.0) & 25.0 \\
\hline
\end{tabular}
\label{tab:domain}
\end{table}

\subsection{Estimation of Spatial Resolution}
\begin{figure}
\centering
\includegraphics[width=.9\textwidth]{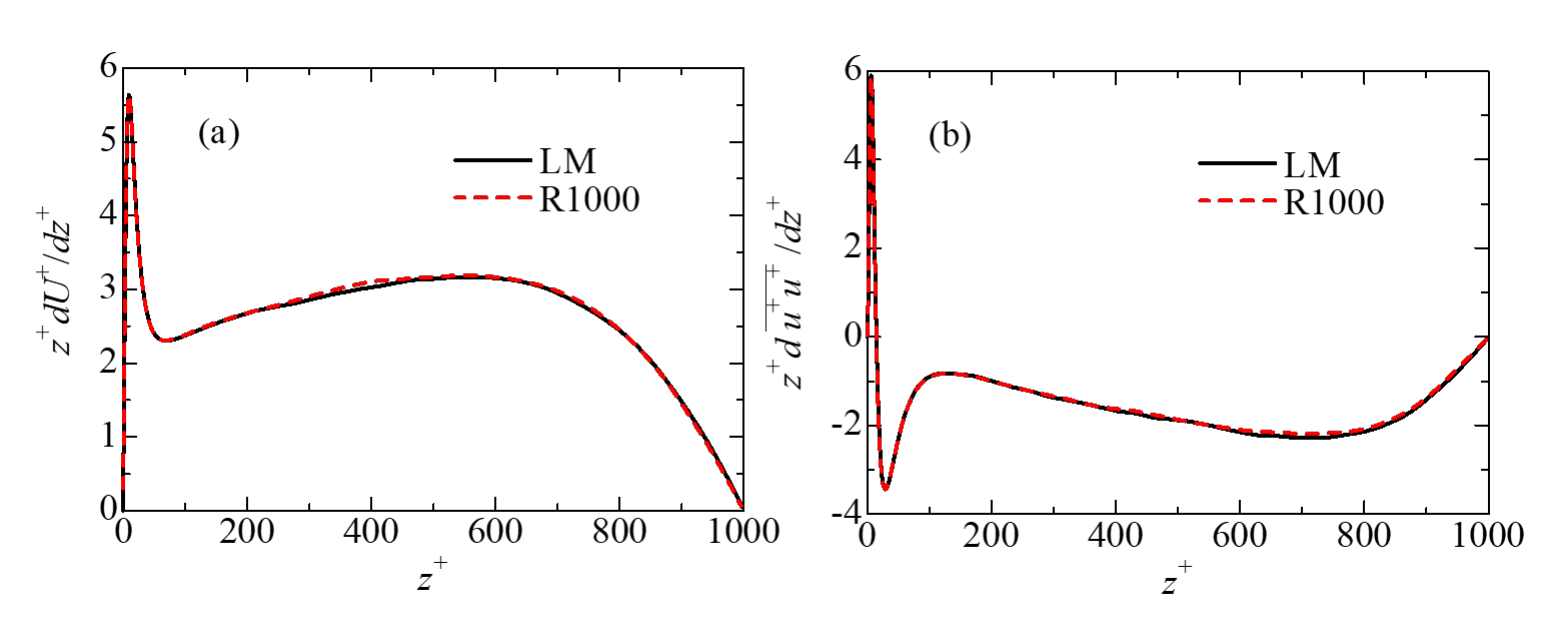}
\caption[ ]{
Comparison between R1000 and LM data\cite{lee2015direct} for pre-multiplied profiles of wall-normal derivative of streamwise mean velocity and turbulent intensities: (a) Streamwise mean velocity, (b) Streamwise normal Reynolds stress.
}
\label{fig:z-dUdz-duudz-R1000}
\end{figure}
\begin{figure}
\centering
\includegraphics[width=.9\textwidth]{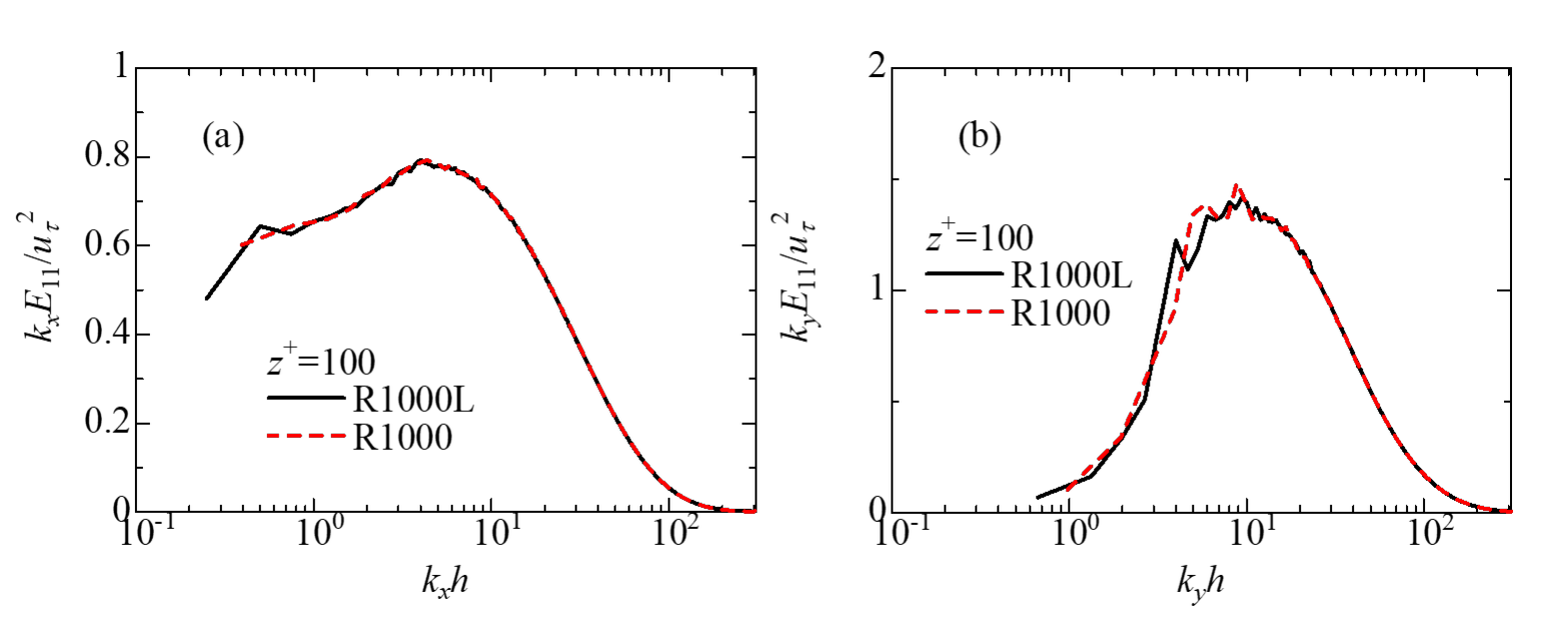}
\caption[ ]{
Comparison between R1000 and R1000L for the pre-multiplied spectra of $u$ at $z^{+}=100$: (a) Streamwise direction, (b)Spanwise direction.
}
\label{fig:PMS-uu-zp100-R1000}
\end{figure}
R1000L, as shown in Table \ref{tab:LM}, ensures the resolution at least twice the Kolmogorov wavenumber in the wall-normal direction $(\Delta z^{+} < \Delta_{2}^{+} = 0.5 \Delta_{1}^{+})$. The resolution in the spanwise direction is approximately equal to the Kolmogorov wavelength $(\Delta y^{+} \approx \Delta_{1}^{+})$, whereas the resolution in the streamwise direction is about half of the Kolmogorov wavelength $(\Delta x^{+} \approx 2\Delta_{1}^{+})$. Therefore, as the required spatial resolution for FTS, the wall-normal direction is set to the same resolution as R1000L, and in the streamwise and spanwise directions, the resolution capable for resolving Kolmogorov wavenumber $(\Delta x^{+} = \Delta y^{+} \approx \Delta_{1}^{+})$ is chosen as the first candidate. The validity of this resolution is confirmed by ensuring that the dissipation spectra in the streamwise and spanwise directions sufficiently decay at the Kolmogorov wavenumber.

\subsection{Estimation of Domain size}
To reduce the computational cost of FTS, we consider reducing the size of the computational domain. In our previous studies\cite{yamamoto2018numerical, kaneda2021velocity}, we applied a computational domain of $L_{x}/h = 16$ and $L_{y}/h = 6.4$, which we now consider for evaluation. R1000 in Table\ref{tab:domain} represents the case where the computational domain is reduced to $L_{x}/h = 16$ and $L_{y}/h = 6.4$, while maintaining the same spatial resolution as R1000L. Figure \ref{fig:z-dUdz-duudz-R1000} presents the comparison between R1000 and LM for wall-normal derivative of the streamwise mean velocity and normal stress, both multiplied by the wall-normal height $z$ . Regardless of the differences in the computational domain size, the two results are almost identical. Figure \ref{fig:PMS-uu-zp100-R1000} shows the comparison between R1000 and R1000L for the pre-multiplied spectrum of $u$ at $z^{+} \approx 100$. Note that the one-dimensional energy spectrum $E_{ii}$ defined as follows:
\begin{equation}
\overline{u_i u_i}
= \int_{2\pi/L_x}^{2\pi/\Delta x} E_{ii}(k_x)\, dk_x
= \int_{2\pi/L_y}^{2\pi/\Delta y} E_{ii}(k_y)\, dk_y .
\label{eq:spect}
\end{equation}
The streamwise turbulence intensity is the physical quantity most affected by the size of the computational domain. However, in the case of this computational domain difference, the variations in spectral distribution are extremely small. As a result, their impact on turbulence statistics is confirmed to be negligible. Based on these observations, in this study's FTS, the grid resolution capable of resolving the Kolmogorov wavenumber is applied to the computational domain with $L_{x}/h = 16$ in the streamwise direction and $L_{y}/h = 6.4$ in the spanwise direction. Meanwhile, in the wall-normal direction, the resolution that can resolve at least twice the Kolmogorov wavenumber, similar to R1000L, is applied. This condition is also shown in Table \ref{tab:domain}.
As demonstrated by the comparisons between R1000L and R1000 in Figs.\ref{fig:z-dUdz-duudz-R1000} and \ref{fig:PMS-uu-zp100-R1000}, the reduction of the computational domain size does not affect turbulence statistics or spectral distributions. Therefore, the domain size of R1000 ($L_{x}/h = 16$ and $L_{y}/h = 6.4$) is adopted as the basis for Full Turbulence Simulation (FTS). Combined with the Kolmogorov-resolving grid resolution defined in Section 3.1, the resulting FTS conditions are summarized in Table \ref{tab:domain}.

\section{\label{sec:level4}Full Turbulence Simulation (FTS) at $Re_{\tau} \approx 1000$}
\begin{figure}
\centering
\includegraphics[width=.8\textwidth]{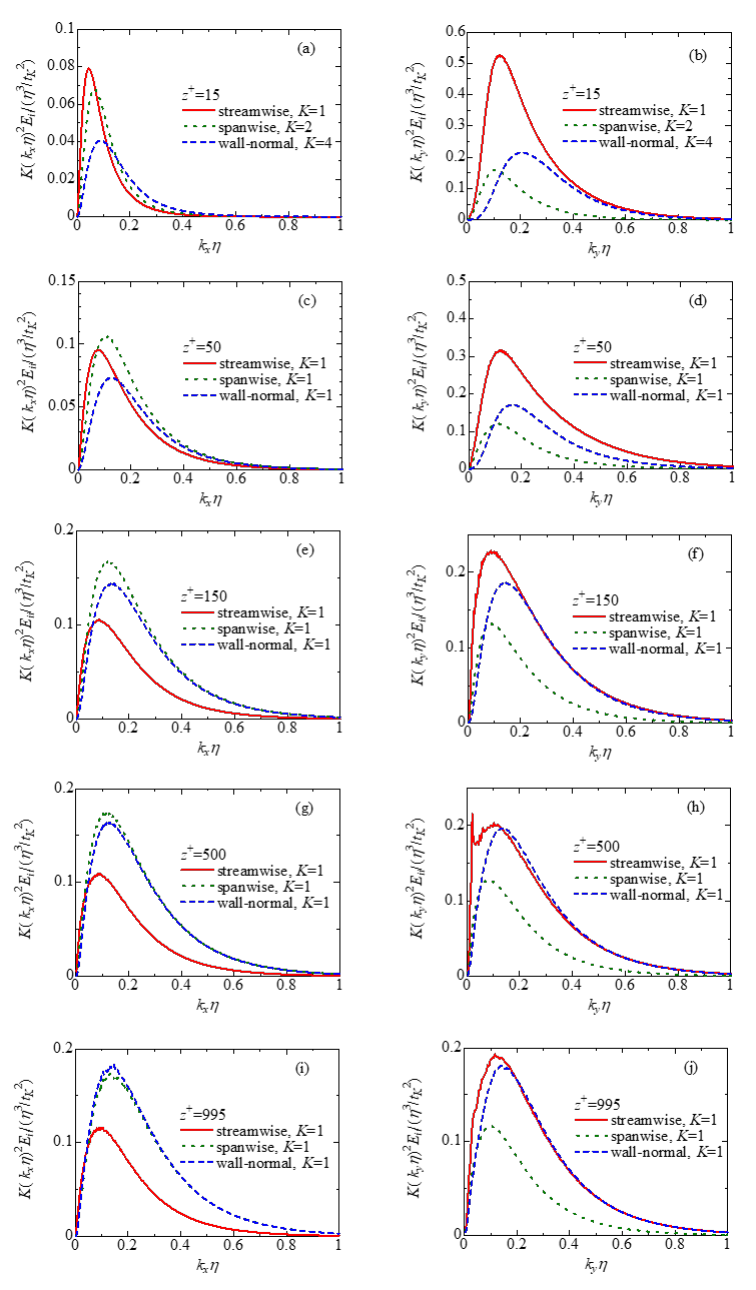}
\caption[ ]{
dissipation spectra, FTS.
}
\label{fig:kkEuvw-xy-FTS}
\end{figure}
\begin{figure}
\centering
\includegraphics[width=.9\textwidth]{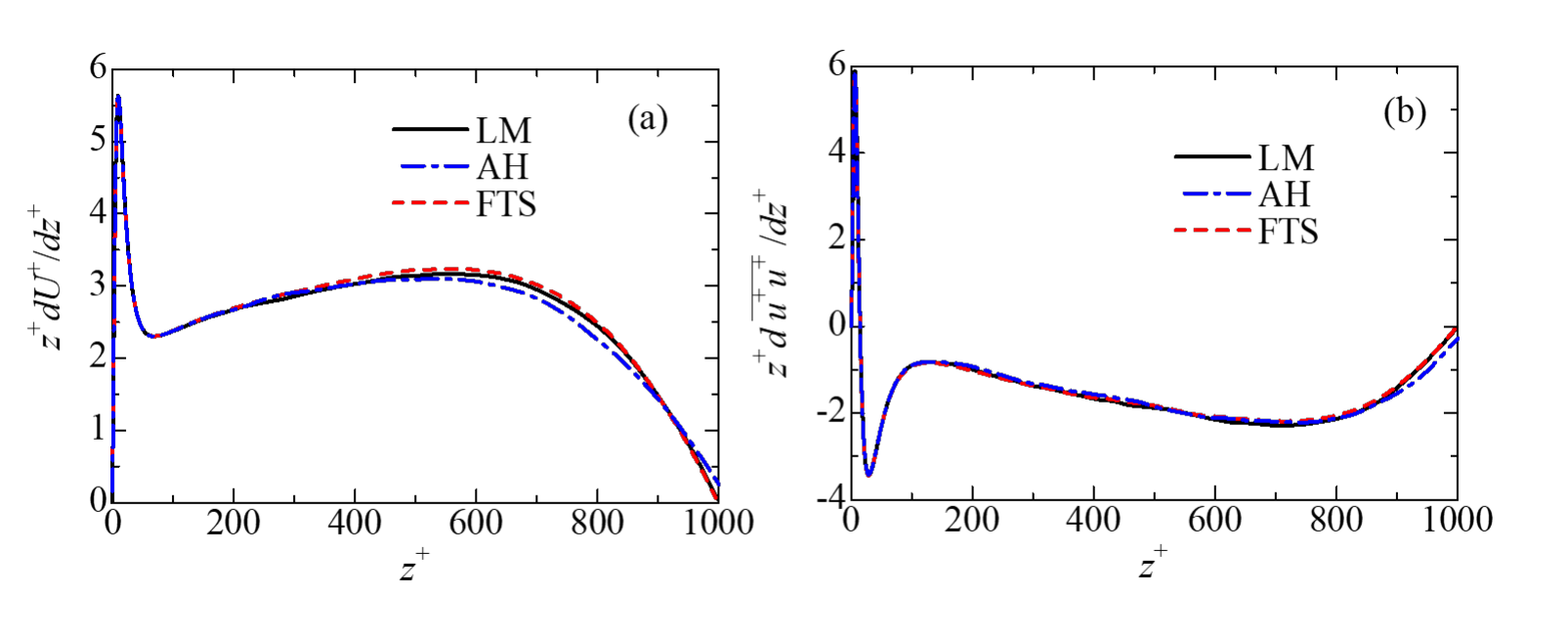}
\caption[ ]{
Comparison of pre-multiplied wall-normal derivatives of the streamwise
mean velocity and streamwise normal Reynolds stress. Results from the present FTS
are compared with LM\cite{lee2015direct}, with AH\cite{alcantara2021direct} included for reference.
}
\label{fig:z-dUdz-duudz-FTS}
\end{figure}
\begin{figure}
\centering
\includegraphics[width=.9\textwidth]{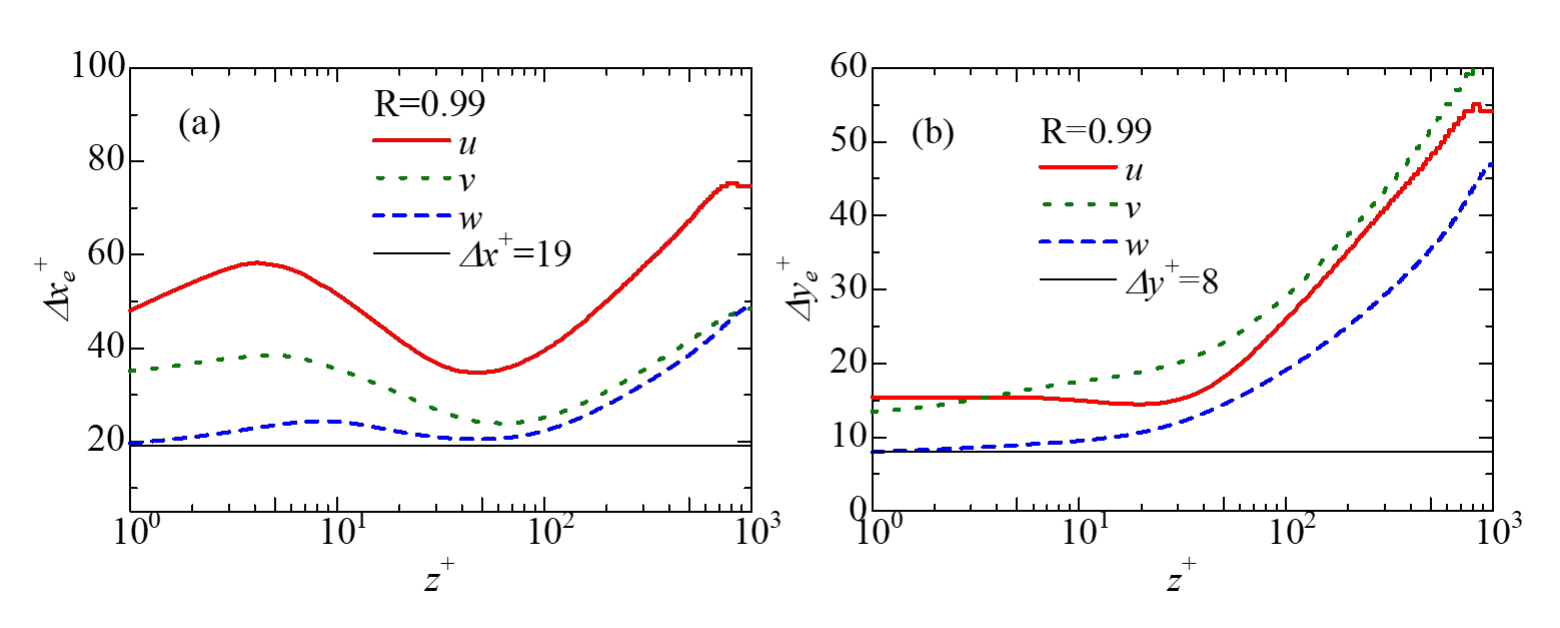}
\caption[ ]{
grid resolution as the spacing necessary to capture 99\% of the kinetic energy, (a) streamwise resolution $(\Delta x_{e}^{+})$, (b) spanwise resolution $(\Delta y_{e}^{+})$.
}
\label{fig:resolution-A}
\end{figure}
\begin{figure}
\centering
\includegraphics[width=.9\textwidth]{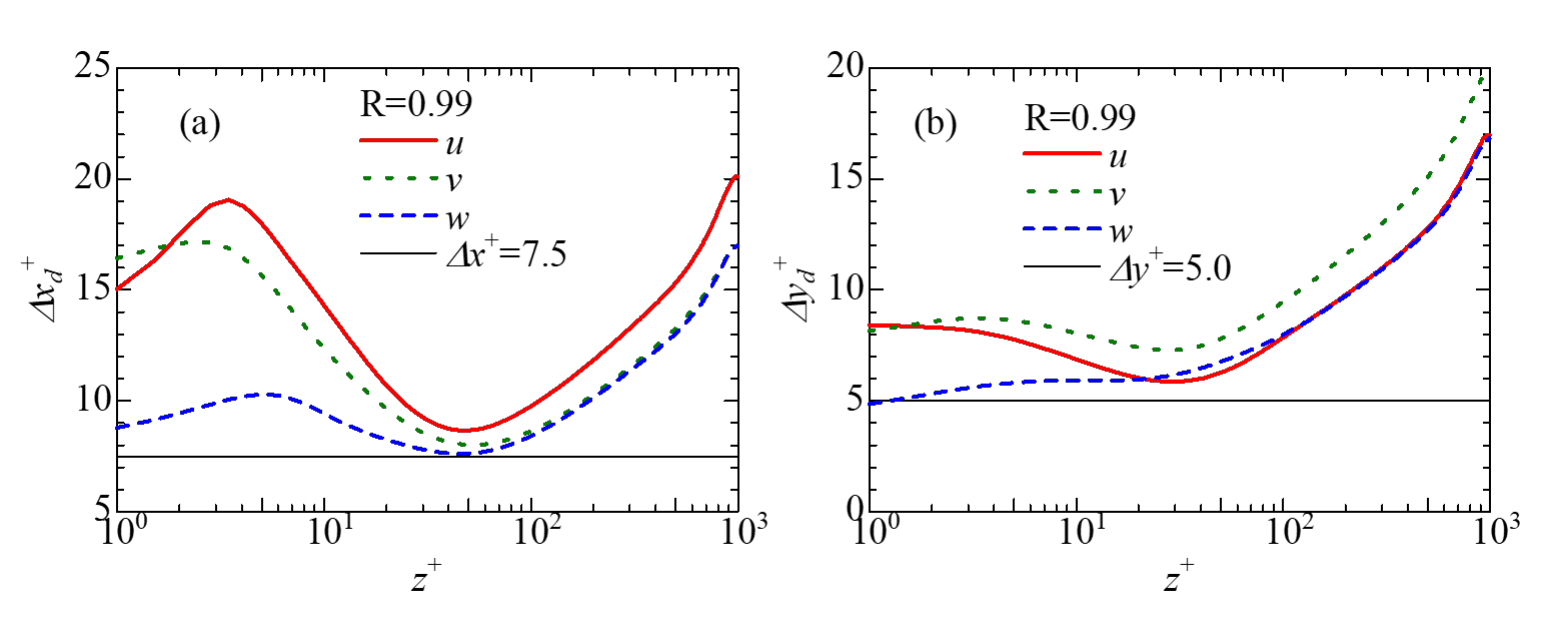}
\caption[ ]{
grid resolution as the spacing necessary to capture 99\% of the dissipation spectra, (a) streamwise resolution $(\Delta x_{d}^{+})$, (b) spanwise resolution $(\Delta y_{d}^{+})$.
}
\label{fig:resolution-DIS}
\end{figure}

\begin{figure}
\centering
\includegraphics[width=.9\textwidth]{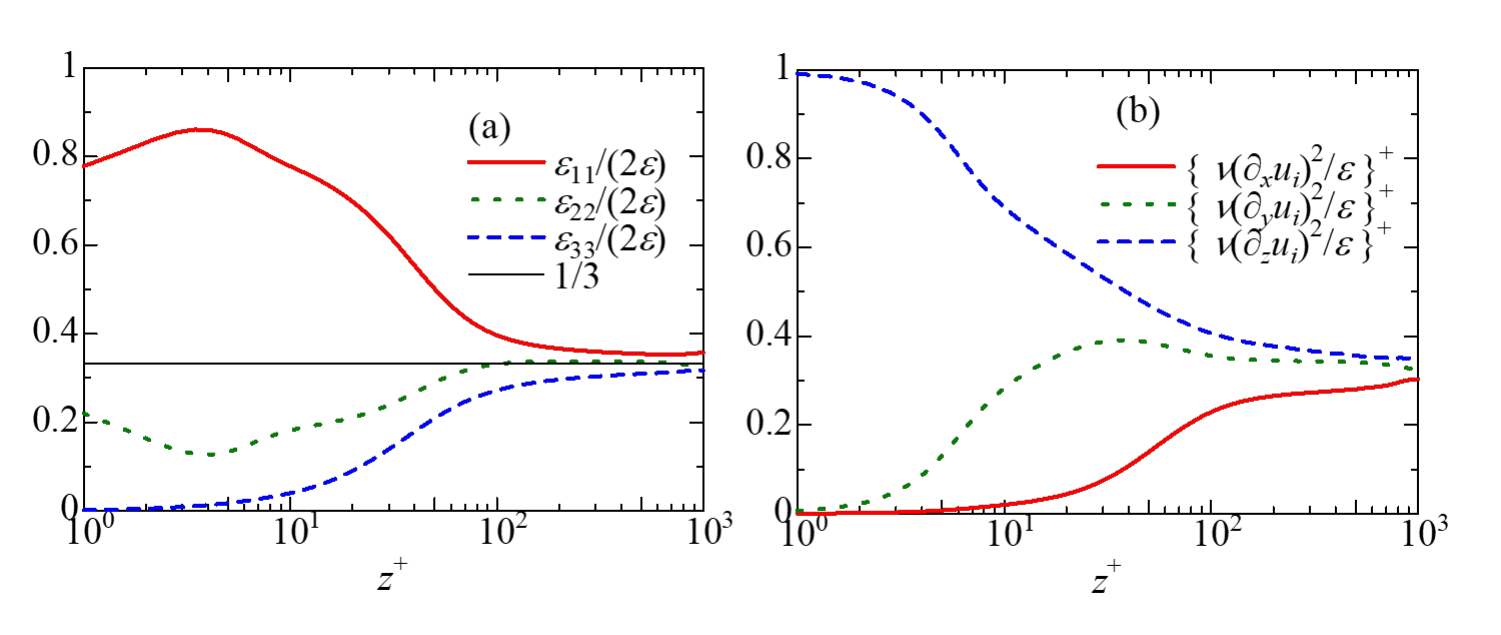}
\caption[ ]{
Anisotropy of dissipation rates: (a) relative contributions of velocity components; (b) relative contributions of derivative directions.
}
\label{fig:dis-FTS}
\end{figure}

\begin{figure}
\centering
\includegraphics[width=.9\textwidth]{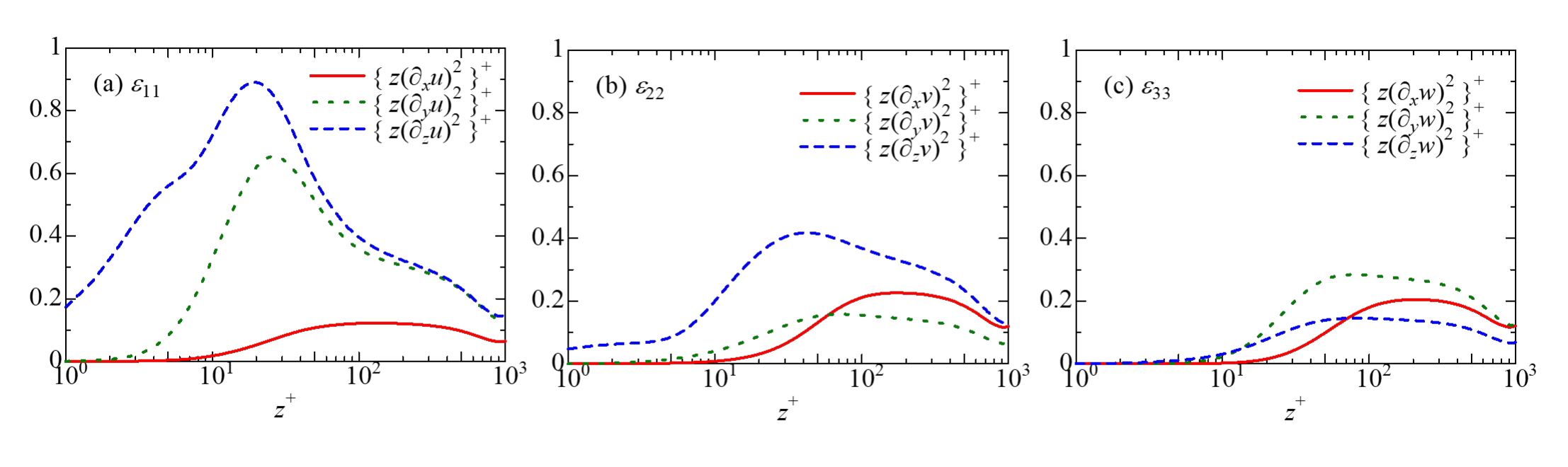}
\caption[ ]{
Directional contributions to dissipation components: (a) $\varepsilon_{11}(u)$, (b)$\varepsilon_{22}$, and (c)$\varepsilon_{33}(w)$.
}
\label{fig:e11-e22-e33}
\end{figure}

Based on the FTS condition shown in Table \ref{tab:domain}, the long-time integration of $T^{+}/Re_{\tau} \approx = 25$ was performed to obtain the turbulent statistics. The computations were carried out on two supercomputing systems: SX-Aurora TSUBASA/32 VEs (512 cores) at the Cyber Science Center, Tohoku University, and FX100/128 nodes (6,144 cores) at the Information Technology Center, Nagoya University. The elapsed time per integration step was approximately 0.5 seconds on both systems, demonstrating the efficiency and scalability of the current DNS code under the FTS condition.

\subsection{Validation of FTS results}
This study validates FTS by demonstrating that the dissipation spectrum decays sufficiently at the Kolmogorov wavenumber. 
Figure \ref{fig:kkEuvw-xy-FTS} presents the streamwise and spanwise dissipation spectra at several wall-normal heights. The spectra are normalized by the Kolmogorov length scale $\eta$ and the Kolmogorov time scale ($t_{K} \equiv (\nu/\varepsilon)^{1/2}$), and a coefficient $K$ is applied to enhance visualization. As shown in Figure \ref{fig:kkEuvw-xy-FTS}, the spectral values for the streamwise, spanwise, and wall-normal components have been sufficiently decayed at the Kolmogorov wavenumber, thereby verifying the validity of this calculation as an FTS. Figure \ref{fig:z-dUdz-duudz-FTS} compares the pre-multiplied wall-normal derivatives of the streamwise mean velocity and streamwise normal Reynolds stress between the present FTS and LM\cite{lee2015direct}, AH\cite{alcantara2021direct}. Both quantities are sensitive to resolution and domain configuration, especially in the near-wall and intermediate regions. The excellent agreement observed across the entire wall-normal range confirms that the present FTS reproduces the turbulence statistics of LM\cite{lee2015direct} with high fidelity. A comparison with AH\cite{alcantara2021direct} shows that, while some differences appear in the outer region of the mean-velocity profile, their results agree well with the present FTS in the near-wall region.These results, combined with the dissipation spectrum validation in Figure \ref{fig:kkEuvw-xy-FTS}, establishes the reliability of the current simulation as a Full Turbulence Simulation.
In this study, Full Turbulence Simulation (FTS) refers to DNS that resolves
the Kolmogorov wavenumber in all spatial directions. The validity of the resolution was confirmed through dissipation spectra and statistical comparisons with established DNS benchmarks, ensuring that the present simulation satisfies the criteria for FTS as defined in Chapter \ref{sec:level3}.

\subsection{First-Approximation DNS Resolution Derived from FTS (Full Energy-Resolution Criterion)}
In this study, we define a first-approximation DNS resolution criterion as the grid spacing required to resolve 99\% of the kinetic energy in each velocity component. This condition is extracted from the conventional energy spectra obtained by FTS, representing the full energy-resolution requirement to capture all turbulent motions. Strictly speaking, the dissipation spectra should also be examined to ensure sufficient decay at the Kolmogorov wavenumber. However, as a first approximation, the full kinetic-energy resolution condition provides a rational estimate of the required grid resolution. 
The required grid resolution, defined as the spacing necessary to capture 99\% of the kinetic energy in each velocity component, is computed from the one-dimensional spectra $E_{ii} (k_{x}, z)$ and $E_{ii} (k_{y}, z)$ obtained via FTS in the streamwise and spanwise directions, respectively, using the following expression:
\begin{equation}
R \overline{u_i u_i(z)}
= \int_{2\pi/L_x}^{2\pi/\Delta x_{e}} E_{ii}(k_{x}, z)\, dk_x
= \int_{2\pi/L_y}^{2\pi/\Delta y_{e}} E_{ii}(k_{y}, z)\, dk_y .
\label{eq:spect-filter}
\end{equation}
Here, $R$ denotes the cumulative energy fraction, with $0 < R \le 1$. The full energy-resolution condition corresponds to $R = 0.99$, and the required grid spacings are defined as $\Delta x_{e}$ and $\Delta y_{e}$.
Figure 9 shows the values of $\Delta x_{e}$ and $\Delta y_{e}$ obtained from Eq. (\ref{eq:spect-filter}) at each wall-normal height $z$. In both the streamwise and spanwise directions, the most stringent requirement is imposed by the wall-normal velocity component $w$, from which the minimal values are estimated as $\Delta x_{e}^{+} \approx 19$ and $\Delta y_{e}^{+} \approx 8$. These resolutions are consistent with those employed in spectral DNS in the wall-parallel directions with relatively coarse grids, such as Kasagi et al.\cite{kasagi1992direct} ($Re_{\tau} \approx 150, \Delta x^{+} \approx 18.4$ and $\Delta y^{+} \approx 7.4$),  Kim \& Hussain\cite{kim1993propagation}($Re_{\tau} \approx 180, \Delta x^{+} \approx 18$ and $\Delta y^{+} \approx 9$), and Kaneda \& Yamamoto\cite{kaneda2021velocity} ($Re_{\tau} \approx 8000, \Delta x^{+} \approx 18.5$ and $\Delta y^{+} \approx 8.9$).

\subsection{\label{subsec:level4.4}DNS with Full Dissipation Resolution Derived from FTS}
In turbulent flows, it is ideal not only to resolve the kinetic energy but also to fully capture the dissipation processes. From this perspective, we define DNS with Full Dissipation Resolution as the grid resolution required to resolve 99\% of the dissipation spectra in each velocity component.
This criterion is computed from the one-dimensional dissipation spectra obtained via FTS, using the following expression:
\begin{equation}
R\overline{\left( \frac{\partial u_i}{\partial x} \right)^2} (z)
=
\int_{2\pi/L_x}^{\pi/\Delta x_d}
k_x^{2}\, E_{ii}(k_x, z)\, dk_x .
\label{eq:dis-spect-filter}
\end{equation}
Equation (\ref{eq:dis-spect-filter}) represents the cumulative dissipation in the streamwise direction. A similar formulation applies to the spanwise direction ($y$), though omitted here for brevity. Based on this definition, the required grid spacings $\Delta x_{d}^{+}$ and $\Delta y_{d}^{+}$ determined to ensure 99\% resolution of the dissipation spectra.
Figure \ref{fig:resolution-DIS} presents the values of $\Delta x_{d}^{+}$ and $\Delta y_{d}^{+}$ computed from Eq. (\ref{eq:dis-spect-filter}) at each wall-normal height $z$. As in the case of kinetic energy resolution, the most stringent requirement in both the streamwise and spanwise directions arises from the wall-normal velocity component $w$, yielding minimum estimates of $\Delta x_{d}^{+} \approx 7.5$ and $\Delta y_{d}^{+}\approx 5.0$. These resolutions are not fully satisfied even in DNS studies considered high-resolution, such as Kim et al.\cite{kim1987turbulence}($Re_{\tau} \approx 180, \Delta x^{+} \approx 11.8$ and $\Delta y^{+} \approx 7.1$) and Lee \& Moser\cite{lee2015direct}($Re_{\tau} \approx 5200, \Delta x^{+} \approx 12.7$ and $\Delta y^{+} \approx 6.4$). According to the authors’ knowledge, spectral DNS in the wall-parallel directions that satisfy the full dissipation-resolution criterion ($\Delta x^{+} \approx 7.5$ and $\Delta y^{+} \approx 5.0$) at $Re_{\tau} \ge 1000$ are limited to Yamamoto \& Kunugi\cite{yamamoto2016mhd}($Re_{\tau} \approx 1000, \Delta x^{+} \approx 6.7$ and $\Delta y^{+} \approx 5.0$), Morishita et al.\cite{morishita2019length}($Re_{\tau} \approx 1280, 2560$, and $5120$, $\Delta x^{+} \approx 7.9$ and $\Delta y^{+} \approx 3.9$),  Kaneda \& Yamamoto\cite{kaneda2021velocity} ($Re_{\tau} \approx 1000, \Delta x^{+} \approx 7.2$ and $\Delta y^{+} \approx 4.2$), Alcántara-Ávila \& Hoyas\cite{alcantara2021direct}($Re_{\tau} \approx 1000$ and $2000, \Delta x^{+} \approx 4.1$ and $\Delta y^{+} \approx 2.5$). It should be noted, however, that the cases by Morishita et al.\cite{morishita2019length} employed an unusually small computational domain $(L_{x}/h = \pi, L_{y}/h = \pi/2)$ despite its fine resolution.

\subsection{Resolution Effects on Dissipation Rate}
In anisotropic turbulence such as channel flow, dissipation rates exhibit anisotropy (see Fig. \ref{fig:dis-FTS}(a)). Here, $\varepsilon_{11}, \varepsilon_{22}, \varepsilon_{33}$ are defined in Eq. ( \ref{eq:dis}) as the dissipation rates of the streamwise, spanwise, and wall‑normal turbulent components.
\begin{equation}
\varepsilon_{ii}
= 2\nu \left[
\left( \frac{\partial u_i}{\partial x} \right)^2
+ \left( \frac{\partial u_i}{\partial y} \right)^2
+ \left( \frac{\partial u_i}{\partial z} \right)^2
\right], \qquad
\varepsilon = \frac{\varepsilon_{ii}}{2}.
\label{eq:dis}
\end{equation} 
Near the wall,  $\varepsilon_{11}$ is dominant while the contribution of  $\varepsilon_{33}$ is small, and farther from the wall $(z^{+} > 100)$ a tendency toward isotropy appears. Therefore, when discussing the effects of resolution on dissipation rate, it is necessary to focus on the dominant dissipation terms and their derivative directions. In fact, as shown in Fig. \ref{fig:dis-FTS}(b), wall‑normal derivatives are dominant near the wall, and spanwise derivatives become significant for $z^{+} > 5$. In contrast, streamwise derivatives contribute little up to about $z^{+} < 40$. Since in this study the wall‑normal direction is assumed to be fully resolved down to the Kolmogorov length scale, the resolution of the spanwise direction is the critical factor, making the evaluation in Fig. \ref{fig:resolution-DIS}(b) essential.
Meanwhile, examining the component-wise breakdown of dissipation rates (see Fig. \ref{fig:e11-e22-e33}) reveals that spanwise derivatives, specifically $\partial u/\partial y$ and $\partial v/\partial y$, dominate $\varepsilon_{11}$ and $\varepsilon_{22}$ near the wall $(z^{+} < 30)$; it is therefore necessary to check the required resolution for these terms. To quantify the resolution required, Fig. \ref{fig:resolution-DIS}(b) shows that resolving 99\% of the dissipation spectra for $u$ and $v$ requires a spanwise resolution of approximately $\Delta y_{d}^{+} \approx 8$. While the streamwise vortex remains the primary carrier of turbulent kinetic energy, the present analysis demonstrates that its dissipation fidelity is governed by the accurate reproduction of small‑scale spanwise fluctuations, specifically  $\partial u/\partial y$ and $\partial v/\partial y$. In addition to the reproduction of wall‑normal variations, which is already ensured in the present study, the spanwise resolution thus becomes the critical factor for capturing the dominant dissipation terms. Therefore, the first-approximation resolution $(\Delta x_{e}^{+} \approx 19$, $\Delta y_{e}^{+} \approx 8)$, originally designed to resolve 99\% of kinetic energy, is also likely sufficient to resolve the dominant dissipation terms. Quantitative reproducibility will be examined in detail in the next section.

\section{\label{sec:level5}Accuracy Verification at First-Approximation Resolution}
\begin{table}[t]
\centering
\scriptsize
\caption{Validation of first-approximation resolution
.}

\begin{tabular}{lccccccccccc}
\hline
RUN & discretization ($x,y$) & discretization ($z$) & aliasing error &
$Re_{\tau}$ & $Re_b$ & $L_x/h$ & $L_y/h$ &
$N_x$ ($\Delta x^+$) & $N_y$ ($\Delta y^+$) & $N_z$ ($\Delta z^+$) & $T^+/Re_{\tau}$ \\
\hline
R1000A &
Fourier & CD2 & aliased &
1000 & 20000 & 16 & 6.4 &
900 (17.8) & 864 (7.4) & 512 (0.6--8.0) & 25.0 \\
R1000B &
Fourier & CD2 & 3/2-rule &
1000 & 20000 & 16 & 6.4 &
864 (18.5) & 800 (8.0) & 512 (0.6--8.0) & 25.0 \\
R1000C &
Fourier & CD2 & aliased &
1000 & 20000 & 16 & 6.4 &
800 (20.0) & 640 (10.0) & 512 (0.6--8.0) & 25.0 \\
R1000D &
Fourier & CD2 & aliased &
1000 & 20000 & 16 & 6.4 &
576 (28.5) & 432 (15.2) & 512 (0.6--8.0) & 25.0 \\
\hline
\end{tabular}
\label{tab:R1000A}
\end{table}

\begin{figure}
\centering
\includegraphics[width=.7\textwidth]{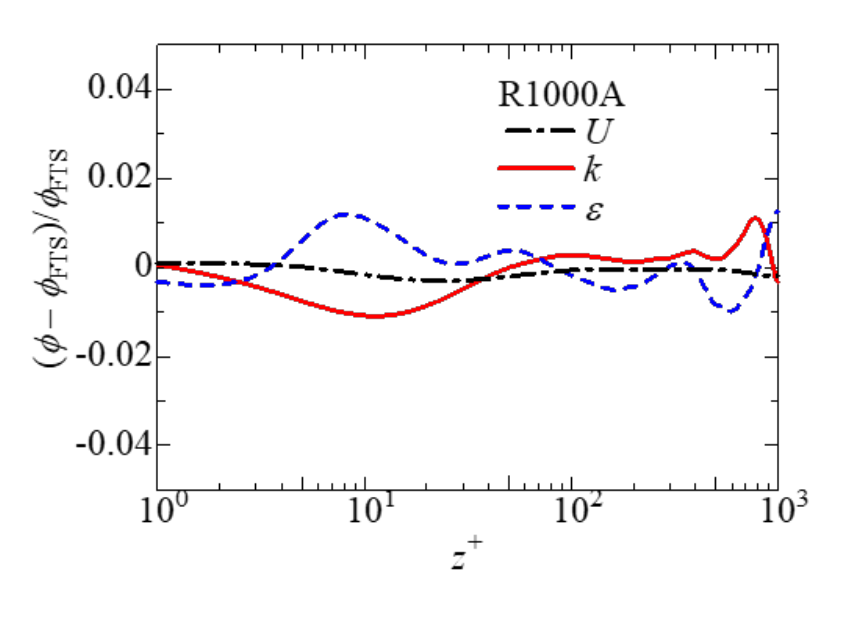}
\caption[ ]{
Relative differences between R1000A and the FTS reference for the mean velocity $U$, turbulent kinetic energy $k$, and dissipation rate $\varepsilon$, plotted as $(\phi - \phi_{FTS})/\phi_{FTS})$ against the wall‑normal coordinate $z^{+}$.
}
\label{fig:error-R1000A}
\end{figure}

\begin{figure}
\centering
\includegraphics[width=.7\textwidth]{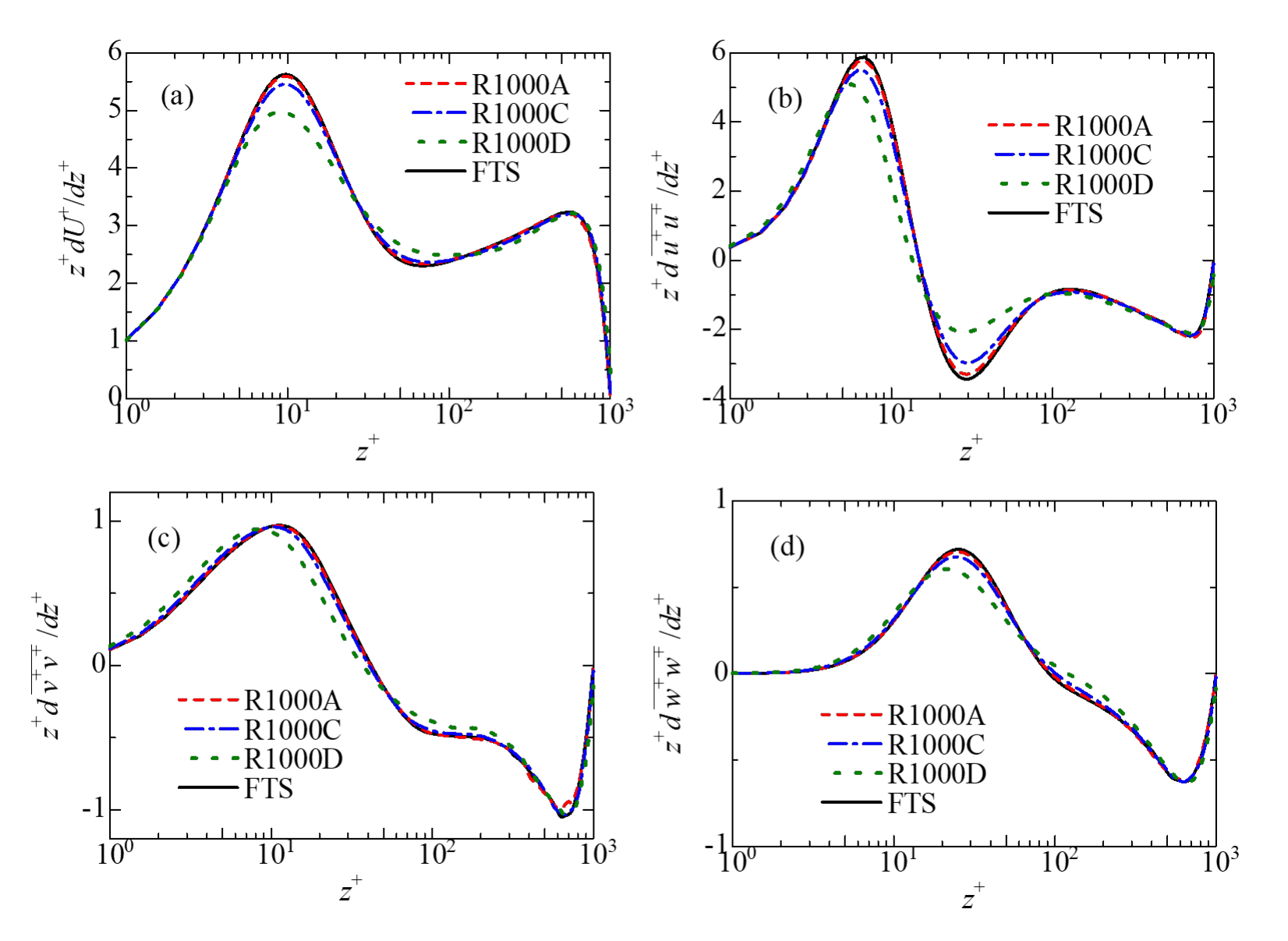}
\caption[ ]{
Effect of spatial resolution on the wall‑normal derivatives of the mean velocity and turbulent normal stresses. The reference FTS data are shown by solid black lines, while R1000A, R1000C, and R1000D are indicated by red, blue, and green dashed lines, respectively. 
}
\label{fig:R1000ACD}
\end{figure}

\begin{figure}
\centering
\includegraphics[width=.7\textwidth]{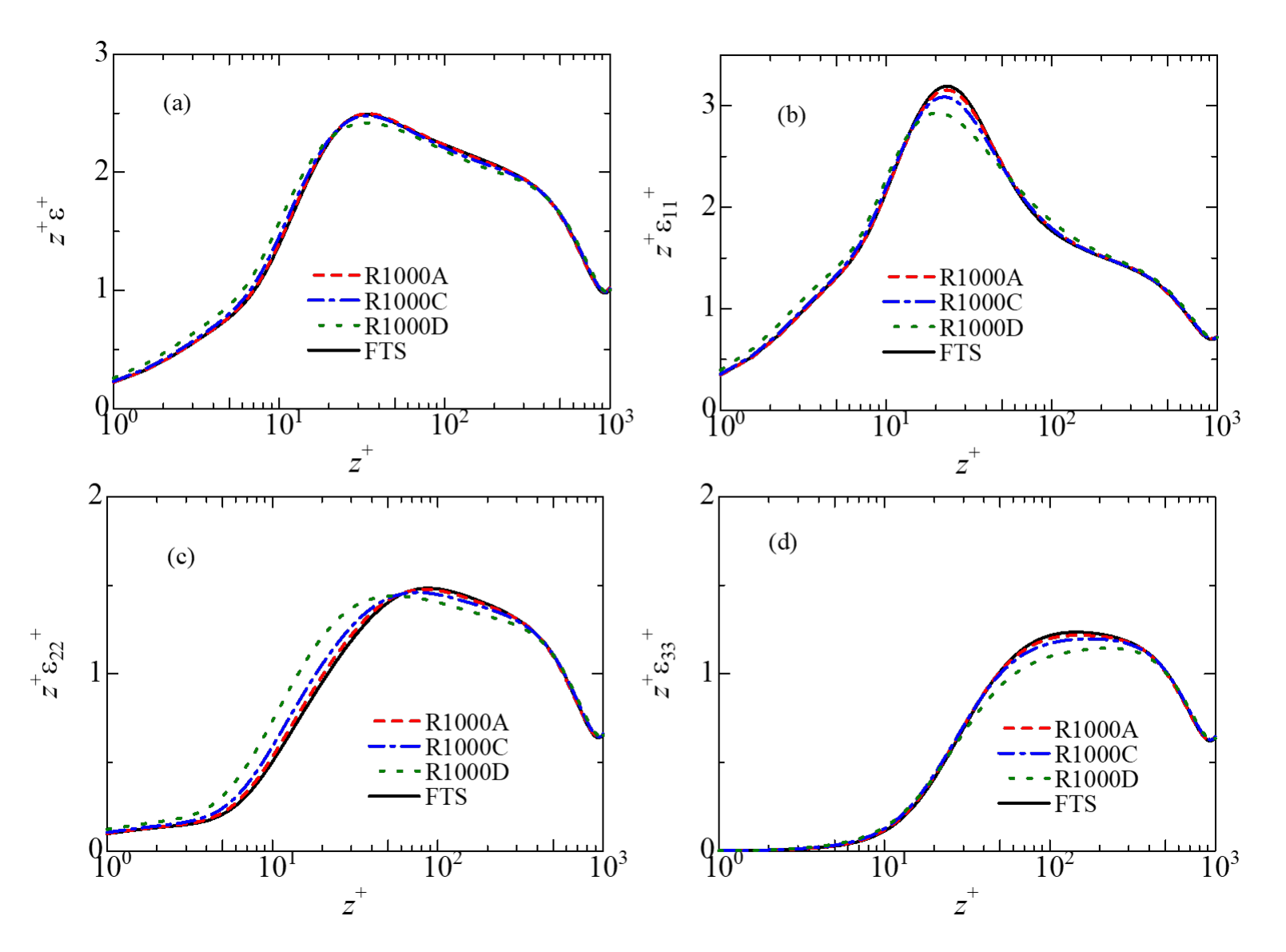}
\caption[ ]{
Effect of spatial resolution on the dissipation rate of turbulent kinetic energy and the dissipation rates of the individual normal‑stress components. The reference FTS data are shown by solid black lines, while R1000A, R1000C, and R1000D are indicated by red, blue, and green dashed lines, respectively.
}
\label{fig:e-R1000ACD}
\end{figure}

\begin{figure}
\centering
\includegraphics[width=.7\textwidth]{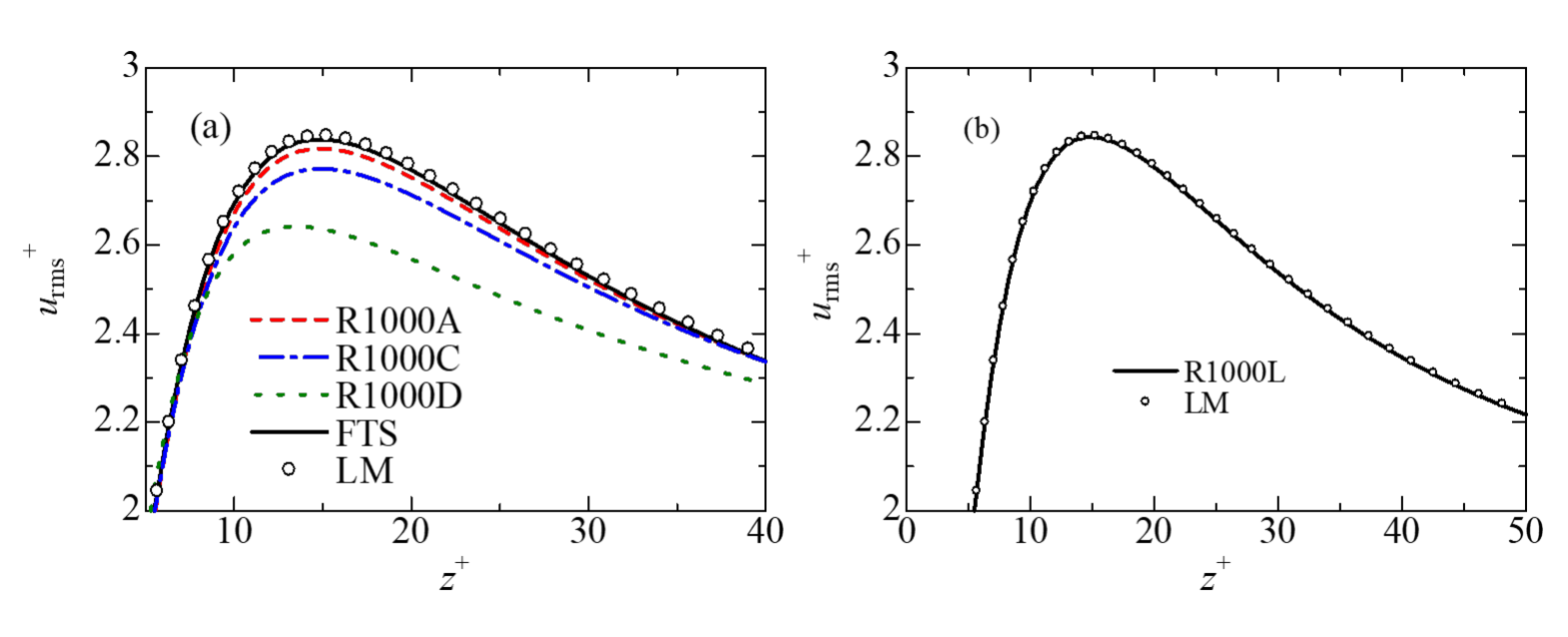}
\caption[ ]{
Effect of spatial resolution and computational domain on the streamwise turbulent intensity $u_{rms}^{+}$.
(a) Comparison among the FTS reference (black solid line), the R1000A, R1000C, and R1000D cases (red, blue, and green dashed lines, respectively), and the LM data (open circles)\cite{lee2015direct}. (b) Comparison between the present R1000L case—computed with the same resolution and computational domain as LM—and the LM data\cite{lee2015direct}.
}
\label{fig:urms-peak-R1000A}
\end{figure}

\begin{figure}
\centering
\includegraphics[width=.7\textwidth]{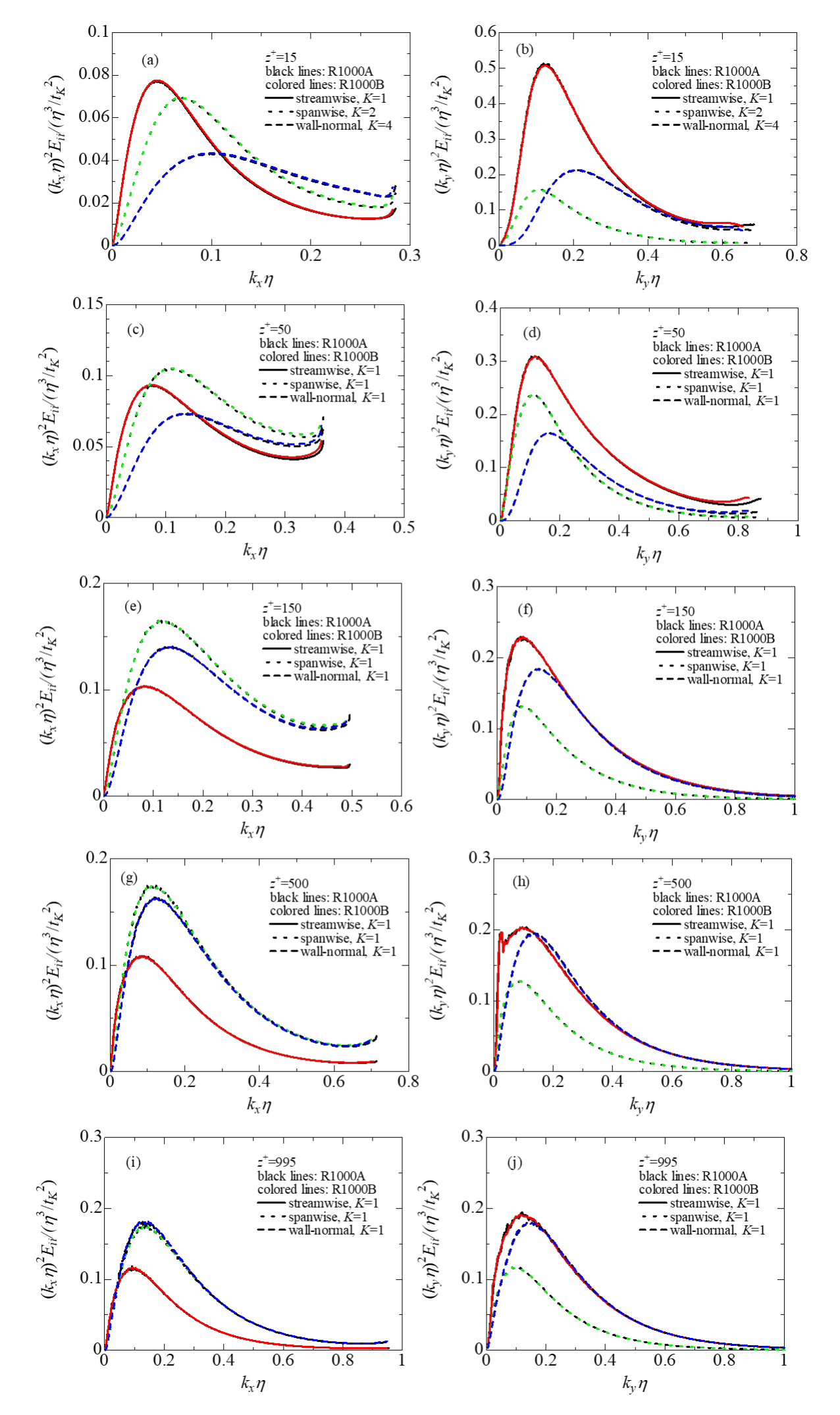}
\caption[ ]{
Comparison of the dissipation spectra of the turbulent normal‑stress components at several wall‑normal locations for R1000A (black lines) and the fully de‑aliased R1000B case (colored lines). Solid, dashed, and dotted lines denote the streamwise, spanwise, and wall‑normal components, respectively.
}
\label{fig:spect-R1000A}
\end{figure}

In the following chapter, we first perform DNS computations at the first-approximation resolution $(\Delta x^{+} \approx 19, \Delta y^{+} \approx 8)$ and verify their accuracy. The objective is to demonstrate that, although this resolution does not fully satisfy the dissipation criterion, it is sufficient to reproduce the essential turbulence statistics. Validation is carried out through comparisons with FTS result and benchmark DNS data reported by Lee \& Moser\cite{lee2015direct}. In addition, the potential influence of aliasing errors, which has been deferred in the previous sections, is explicitly examined here to confirm that the present resolution does not introduce spurious spectral distortions.
\subsection{Numerical conditions}
Table \ref{tab:R1000A} summarizes the validation conditions at the first-approximation resolution. In the present code, the effect of the high-wavenumber cutoff filter defined in Eq. (\ref{eq:filter}) (62/64, corresponding to approximately 97\% of the effective resolution) was taken into account to set the R1000A condition. In addition, a companion run R1000B was performed with aliasing errors completely removed by the 3/2 rule, providing a direct comparison of the influence of aliasing on turbulence statistics. Note that aliasing errors generally become problematic only when the resolution is insufficient. In fact, under the conditions of Tables \ref{tab:LM} and \ref{tab:domain}, their influence is undetectable not only in turbulence statistics but also in spectral distributions. At the present resolution (corresponding to 99\% of the kinetic energy coverage), however, aliasing errors may potentially appear in the dissipation spectra, and this possibility is explicitly examined in the following validation. Furthermore, to clarify the influence of spatial resolution and the contribution of aliasing errors to numerical stability, additional simulations—R1000C and R1000D—were conducted by intentionally degrading the resolution from the R1000A baseline$(\Delta x^{+} \approx 17.8, \Delta y^{+} \approx 7.4)$. These cases were designed to fall below the first‑approximation DNS resolution, allowing us to systematically evaluate how insufficient resolution affects turbulence statistics, dissipation spectra, and the emergence of spurious high‑wavenumber energy associated with aliasing. By comparing these degraded‑resolution runs with R1000A and the fully de‑aliased R1000B, we identify the threshold at which aliasing errors begin to manifest and quantify their impact on numerical stability and accuracy.

\subsection{Validation of First-Approximation Resolution}
To quantitatively verify that the first‑approximation resolution of R1000A indeed reproduces the fully resolved statistics, we compare the mean velocity $U$, turbulent kinetic energy $k=( \overline{uu} +\overline{vv} + \overline{ww} )/2$, and dissipation rate $\varepsilon$ with the FTS reference. Figure \ref{fig:error-R1000A} shows the relative differences $(\phi - \phi_{FTS})/\phi_{FTS}$. Here, $\phi$ and $\phi_{FTS}$ represent the results obtained using of R1000A and FTS, respectively. Across the entire wall‑normal range, the deviations remain within 1\% for all three quantities. This confirms that the 99\% energy‑resolution criterion derived from the FTS spectra is sufficient to reproduce not only the mean flow but also the turbulence energy and dissipation statistics with high fidelity.
Figure \ref{fig:R1000ACD} illustrates the influence of spatial resolution on the wall‑normal derivatives of the mean velocity and turbulent normal stresses. The R1000A case shows virtually perfect agreement with the FTS results, with no discernible differences at the visual level. When the resolution is slightly degraded, as in R1000C, small but noticeable deviations begin to appear, including a mild underprediction of peak values. In contrast, the low‑resolution case R1000D exhibits a clearly pronounced underestimation of the peak magnitudes, indicating a substantial deterioration in accuracy.
Figure \ref{fig:e-R1000ACD} presents the influence of spatial resolution on the dissipation rate of turbulent kinetic energy and on the dissipation rates of the individual normal stress components, as defined in Eq. (\ref{eq:dis}). Similar to the trends observed for the mean quantities and turbulent normal stresses in Figure  \ref{fig:e-R1000ACD}, the R1000A case exhibits virtually perfect agreement with the FTS results, with no visually discernible differences. As the resolution is degraded, however, discrepancies become increasingly evident: R1000C shows a mild underprediction of peak values, while the low‑resolution case R1000D exhibits a clearly pronounced underestimation across all components. Although not shown here, the R1000A case also matches the FTS results for all terms in the transport equations of the turbulent normal stresses, further confirming the adequacy of the present resolution. This behavior is consistent with the analysis in Chapter \ref{sec:level4}, where it was demonstrated that R1000A resolves not only the kinetic‑energy‑containing motions but also the dominant contributions to the dissipation terms, thereby ensuring sufficient accuracy.

It is well known that, even at resolutions comparable to R1000A, the peak value of the streamwise turbulent intensity tends to be slightly underpredicted (see, e.g., \cite{kaneda2021velocity}). Figure \ref{fig:urms-peak-R1000A}(a) shows the effect of spatial resolution on the streamwise turbulent intensity. A modest underestimation of the peak is observed in R1000A, and this discrepancy becomes more pronounced as the resolution is further degraded, as seen in R1000C and R1000D. The cause of this behavior is unlikely to be a modification of the turbulence‑transport mechanisms, since all terms in the transport equation of the turbulent normal stresses for R1000A agree closely with the FTS results. Rather, this trend is attributed to the structure of wall‑bounded turbulence: production occurs exclusively in the streamwise component, and the majority of this production is concentrated near the wall at the location of the intensity peak. Consequently, fluctuations below the R1000A resolution scale contribute to the peak magnitude, and their partial omission manifests as the observed underprediction.
In addition to the resolution effects discussed above, Figure \ref{fig:urms-peak-R1000A}(b) compares the present R1000L case—computed with the same resolution and computational domain as LM—with the LM data. The two results agree almost perfectly, including the near‑wall peak of $u_{rms}$. This confirms that the numerical scheme used in the present study does not underestimate the peak when the resolution and domain match those of LM.
The difference between FTS and R1000A (FTS $>$ R1000A) therefore reflects the contribution of high‑wavenumber energy that becomes visible only at the FTS resolution. In contrast, the difference between LM and FTS (LM $>$ FTS) is likely related to the smaller computational domain used in the FTS, as suggested by the agreement between R1000L and LM. Although the precise sensitivity of the peak of $u_{rms}$ to domain size and resolution is not fully established, these observations indicate that its interpretation requires caution. A more detailed discussion of these resolution‑dependent effects is presented in the next section.
It should be emphasized that this interpretation applies specifically to the R1000A case. Since all terms in the transport equations of the turbulent normal stresses in R1000A agree closely with the FTS results, the slight underprediction of the streamwise‑intensity peak cannot be attributed to any modification of the turbulence‑transport mechanisms. In contrast, the lower‑resolution cases R1000C and R1000D may exhibit deviations in the transport balance itself, and therefore are not considered within the scope of this interpretation. The behavior observed in R1000A is instead consistent with the structure of wall‑bounded turbulence: production occurs exclusively in the streamwise component, and the majority of this production is concentrated near the wall at the location of the intensity peak. Consequently, fluctuations at scales smaller than those resolved by R1000A contribute to the peak magnitude, and their partial omission manifests as the observed underprediction. 
Figure \ref{fig:spect-R1000A} compares the dissipation spectra in the streamwise and spanwise directions at several wall‑normal locations for R1000A (with aliasing, effective resolutions $\Delta x^{+} \approx 18.4, \Delta y^{+} \approx 7.6$) and R1000B (with aliasing completely removed, effective resolutions, $\Delta x^{+} \approx 18.5, \Delta y^{+} \approx 8.0$). The two cases exhibit nearly perfect agreement up to the highest resolved wavenumbers. In general, when aliasing errors are present under insufficient resolution, the spectra tend to show spurious growth near the maximum wavenumber, often accompanied by numerical instability. The present results demonstrate that the use of the skew‑symmetric form effectively suppresses aliasing errors to a negligible level, even without explicit de‑aliasing. Furthermore, owing to the stabilizing effect of the skew‑symmetric formulation, we confirmed that numerical solutions remain stable not only for R1000D, which employs a reduced resolution, but also for an even more severely under-resolved case ($\Delta x^{+} \approx 100, \Delta y^{+} \approx 50$; figure omitted), without exhibiting numerical instability.
An important advantage of the present aliased formulation is the reduction in both memory usage and computational cost. Compared with the fully de‑aliased R1000B case employing the 3/2 rule, the R1000A configuration achieves approximately a 25\% reduction in memory consumption and about a 10\% reduction in computational cost. Considering applications to ultra‑high‑Reynolds‑number DNS, where memory and communication overhead dominate, these savings are expected to have a substantial impact on overall feasibility and performance.
Yang et al. \cite{yang2021grid} reported that, at increasing Reynolds numbers, insufficient spatial resolution can lead to a noticeable degradation in the prediction of wall‑shear stress due to the under‑resolution of rare, high‑intensity events in the tail of the wall‑stress PDF. This issue is not relevant in the present study, since the fully resolved reference simulation yields $Re_{\tau} = 999$, differing from the target value of 1000 by only 0.1\% (Table \ref{tab:R1000A}).
Taken together, the present validation demonstrates that the spatial resolution of R1000A is fully adequate for DNS at $Re_{\tau} =1000$. The case reproduces the turbulence statistics, dissipation behavior, and transport‑equation balances with high fidelity, while aliasing errors remain negligible owing to the skew‑symmetric formulation. In addition to its accuracy, R1000A offers substantial practical advantages: compared with the fully de‑aliased 3/2‑rule implementation (R1000B), it reduces memory usage by approximately 25\% and computational cost by about 10\%. These combined benefits indicate that R1000A lies on the sufficient‑resolution side of the threshold identified by the FTS and provides a robust and efficient configuration for extending DNS to ultra‑high‑Reynolds‑number regimes. 
At the same time, the Reynolds‑number dependence of the near‑wall peak of the streamwise turbulent intensity has been the subject of active discussion in recent studies, including Marusic et al. \cite{marusic2017scaling}, Cheng \& Sreenivasan\cite{chen2022law}, Cheng \& Fu\cite{cheng2023scale} and Hwang\cite{hwang2024near}. Because these analyses rely sensitively on the accurate representation of near‑wall production and dissipation, it is important to note that R1000A exhibits a slight underprediction of the peak due to its finite streamwise resolution. Consequently, DNS datasets obtained at resolutions comparable to R1000A should be used with caution when interpreting the Reynolds‑number dependence of the streamwise‑intensity peak or when comparing with theoretical predictions. Fully resolved datasets, such as the present FTS, are essential for drawing reliable conclusions regarding the true $Re_{\tau}$ scaling of the near‑wall turbulence intensity.

\section{\label{sec:level6}Resolution Requirement for Capturing the Near‑Wall Turbulence Intensity Peak}
\begin{table}[t]
\centering
\scriptsize
\caption{Resolution conditions used to determine the spatial resolution required to reproduce the near‑wall peak of $u_{rms}$.
}

\begin{tabular}{lccccccccccc}
\hline
RUN & discretization ($x,y$) & discretization ($z$) & aliasing error &
$Re_{\tau}$ & $Re_b$ & $L_x/h$ & $L_y/h$ &
$N_x$ ($\Delta x^+$) & $N_y$ ($\Delta y^+$) & $N_z$ ($\Delta z^+$) & $T^+/Re_{\tau}$ \\
\hline
FTS &
Fourier & CD2 & aliased &
999 & 20000 & 16 & 6.4 &
4000 (4.0) & 1600 (4.0) & 512 (0.6--8.0) & 25.0 \\
R1000 &
Fourier & CD2 & 3/2-rule &
1000 & 20000 & 16 & 6.4 &
1600 (10.0) & 1296 (4.9) & 512 (0.6--8.0) & 25.0 \\
R1000E &
Fourier & CD2 & aliased &
1000 & 20000 & 16 & 6.4 &
1200 (13.3) & 960 (6.7) & 512 (0.6--8.0) & 25.0 \\
R1000F &
Fourier & CD2 & aliased &
1000 & 20000 & 16 & 6.4 &
1152 (13.9) & 960 (6.7) & 512 (0.6--8.0) & 25.0 \\
R1000G &
Fourier & CD2 & aliased &
1000 & 20000 & 16 & 6.4 &
960 (16.7) & 900 (7.1) & 512 (0.6--8.0) & 25.0 \\
R1000A &
Fourier & CD2 & aliased &
1000 & 20000 & 16 & 6.4 &
900 (17.8) & 864 (7.4) & 512 (0.6--8.0) & 25.0 \\
\hline
\end{tabular}
\label{tab:urms}
\end{table}

\begin{figure}
\centering
\includegraphics[width=.7\textwidth]{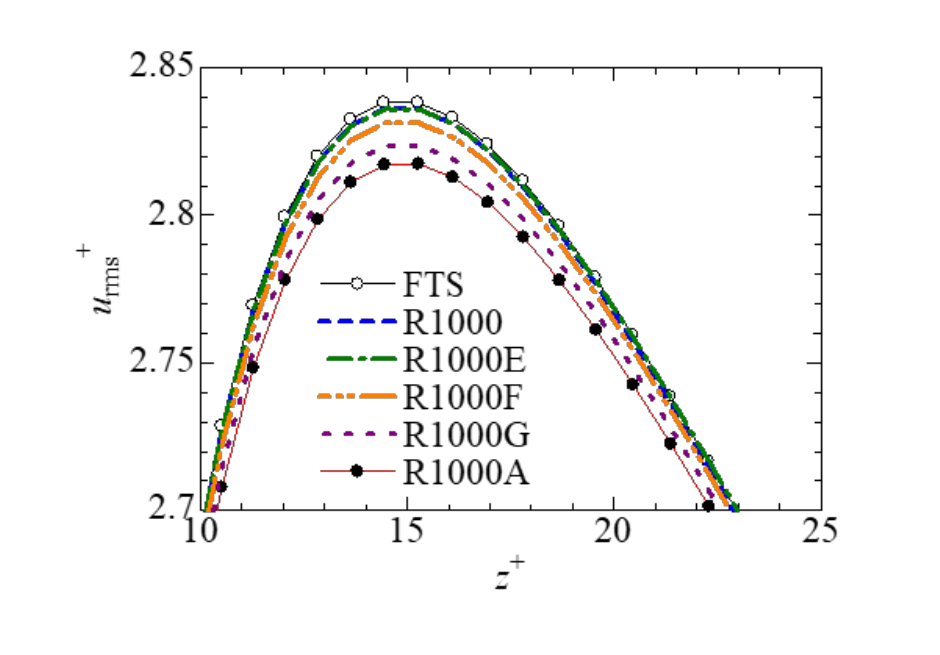}
\caption[ ]{
Near‑wall distribution of the streamwise turbulence intensity $u_{rms}$ for the resolution‑refinement cases listed in Table \ref{tab:urms}.
}
\label{fig:urms-peak-high}
\end{figure}
In Chapter \ref{sec:level5}, the first‑approximation resolution was shown to be sufficient for capturing the majority of the turbulent kinetic energy as well as the dominant contributions to the dissipation terms. At this resolution, the turbulence‑transport mechanisms are faithfully reproduced and remain consistent with the FTS results. However, a slight underprediction of the near‑wall peak of $u_{rms}$ persists, which is attributed to the small amount of turbulent energy residing below the cutoff wavenumber. To quantify the resolution required to reproduce the peak intensity itself, the present section conducts a series of numerical experiments in which the spatial resolution is systematically increased from the R1000A configuration and the convergence of the peak value is examined. Although not discussed in detail here, any resolution that successfully captures the peak intensity naturally satisfies the transport‑mechanism consistency demonstrated in Chapter \ref{sec:level5}.

Table \ref{tab:urms} summarizes the resolution conditions used to determine the spatial resolution required to reproduce the near‑wall peak of $u_{rms}$. The R1000 case corresponds to the baseline resolution introduced in Section 2.3, which is comparable to the LM dataset\cite{lee2015direct}. The R1000E case adopts a resolution similar to that of Kim et al.\cite{kim1987turbulence}, while R1000F and R1000G represent intermediate cases in which the resolution is gradually increased from that of R1000A. These cases allow a systematic assessment of the convergence of the near‑wall turbulence‑intensity peak as the streamwise and spanwise resolutions are refined.
Figure \ref{fig:urms-peak-high} shows the near‑wall distribution of the streamwise turbulence intensity $u_{rms}$ for the resolution‑refinement cases listed in Table \ref{tab:urms}. As the spatial resolution is increased from R1000A toward finer grids, the peak value of $u_{rms}$ gradually rises, indicating that a small amount of turbulent energy residing below the cutoff wavenumber is recovered. The R1000E case, which adopts a resolution comparable to that of Kim et al.\cite{kim1987turbulence}, exhibits near‑complete convergence to the FTS peak. This result suggests that the resolution required to fully capture the near‑wall turbulence‑intensity peak is approximately $\Delta x^{+} \approx 13.3$ and $\Delta y^{+} \approx 6.7$, at which the unresolved fluctuations below the cutoff wavenumber become negligible.
The above results indicate that the spatial resolution required to reproduce the near‑wall peak of $u_{rms}$ is approximately $\Delta x^{+} \approx 13.3$ and $\Delta y^{+} \approx 6.7$. Compared with the first‑approximation resolution used in R1000A, this corresponds to an increase of roughly 1.5 times in the number of grid points in the homogeneous directions. When the reduction of the allowable time‑step size is also taken into account, the total computational cost becomes approximately twice that of R1000A. This quantifies the additional cost associated with resolving the near‑wall turbulence‑intensity peak, and provides a practical guideline for selecting DNS resolutions depending on the required level of accuracy.
As shown in Table \ref{tab:urms}, all cases within the present resolution‑variation study reproduce the wall‑shear stress to within 0.1\% of the FTS value. This includes the R1000 case with $\Delta x^{+} \approx 10.0$ and $\Delta y^{+} \approx 4.9$, indicating that the wall‑shear‑stress prediction is remarkably insensitive to the streamwise and spanwise resolutions considered here. However, this insensitivity also implies that agreement of $Re_{\tau}$ at the 0.1\% level does not guarantee that the flow is resolved with the same fidelity as the FTS. In particular, the dissipation‑spectrum analysis in Section \ref{subsec:level4.4} indicates that resolving approximately 99\% of the dissipation requires grid spacings of $\Delta x^{+} \approx 7.5$ and $\Delta y^{+} \approx 5.0$, whereas the FTS employs even finer spacings of $\Delta x^{+} \approx 4.0$ and $\Delta y^{+} \approx 4.0$. Therefore, if one demands that the wall‑shear‑stress prediction remain accurate without relying on the fortuitous insensitivity observed here, a grid resolution on the order of $\Delta x^{+} \approx 7.5$ and $\Delta y^{+} \approx 5.0$ may be required.
In summary, while the present resolution range is sufficient for reproducing the mean wall‑shear stress, matching the full fidelity of the FTS—particularly for dissipation‑related quantities and the near‑wall peak of $u_{rms}$—requires a finer grid consistent with the dissipation‑resolution criterion.
It is also worth noting that the LM dataset, which is often used as a reference for the near‑wall turbulence‑intensity peak, does not satisfy the dissipation‑resolution criterion discussed above. As shown in Table \ref{tab:urms}, the LM‑equivalent resolution ($\Delta x^{+} \approx 10$ and $\Delta y^{+} \approx 5$) is coarser than the resolution required to resolve approximately 99\% of the dissipation. This implies that the LM peak does not necessarily represent a fully converged value. Indeed, when the present numerical scheme is applied with the same resolution and computational domain as LM (the R1000L case), the peak is reproduced almost exactly, whereas the FTS—despite its higher resolution—yields a slightly lower peak due to its smaller computational domain. If the FTS resolution were applied in the LM domain, the peak would likely exceed the LM value. These observations suggest that the LM peak should not be regarded as a universally robust benchmark for $u_{rms}$, and that its interpretation requires the same caution as the resolution‑dependent behavior examined in this section.

\section{\label{sec:level7}Estimation of the Equivalent Second‑Order Central Difference Resolution Corresponding to the First‑Approximation Accuracy}
\begin{table}[t]
\centering
\scriptsize
\caption{Resolution conditions used to determine the spatial resolution required to reproduce the near‑wall peak of $u_{rms}$.
}

\begin{tabular}{lccccccccccc}
\hline
RUN & discretization ($x,y$) & discretization ($z$) & aliasing error &
$Re_{\tau}$ & $Re_b$ & $L_x/h$ & $L_y/h$ &
$N_x$ ($\Delta x^+$) & $N_y$ ($\Delta y^+$) & $N_z$ ($\Delta z^+$) & $T^+/Re_{\tau}$ \\
\hline
R1000A &
Fourier & CD2 & aliased &
1000 & 20000 & 16 & 6.4 &
900 (17.8) & 864 (7.4) & 512 (0.6--8.0) & 25.0 \\
CD2-1 &
CD2 & CD2 & NA &
994 & 20000 & 16 & 6.4 &
1600 (9.9) & 1296 (4.9) & 512 (0.6--8.0) & 25.0 \\
CD2-2 &
CD2 & CD2 & NA &
993 & 20000 & 16 & 6.4 &
2048 (7.8) & 1350 (4.7) & 512 (0.6--8.0) & 25.0 \\
CD2-3 &
CD2 & CD2 & NA &
994 & 20000 & 16 & 6.4 &
3200 (5.0) & 1400 (4.5) & 512 (0.6--8.0) & 25.0 \\
\hline
\end{tabular}
\label{tab:CD2}
\end{table}
\begin{figure}
\centering
\includegraphics[width=.7\textwidth]{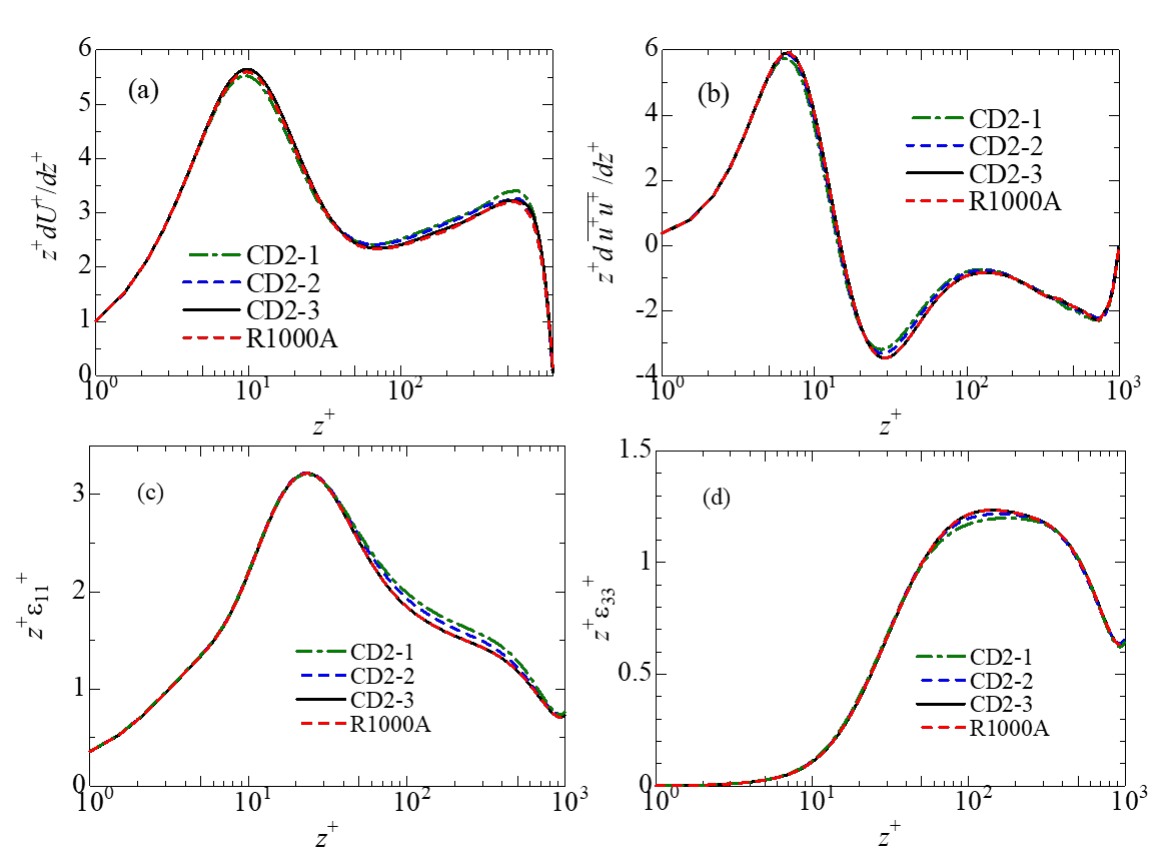}
\caption[ ]{
Comparison of (a) the mean velocity gradient $z^{+}dU^{+}/dz^{+}$, (b) the streamwise turbulence intensity gradient $z^{+}d\overline{u^{+}u^{+} }/dz^{+}$, (c) the dissipation rate of the streamwise turbulence intensity $z^{+} \varepsilon_{11}^{+}$ , and (d) the dissipation rate of the wall‑normal turbulence intensity $z^{+} \varepsilon_{33}^{+}$ for the CD2 cases and the reference spectral solution R1000A.
}
\label{fig:CD2}
\end{figure}

\begin{figure}
\centering
\includegraphics[width=.7\textwidth]{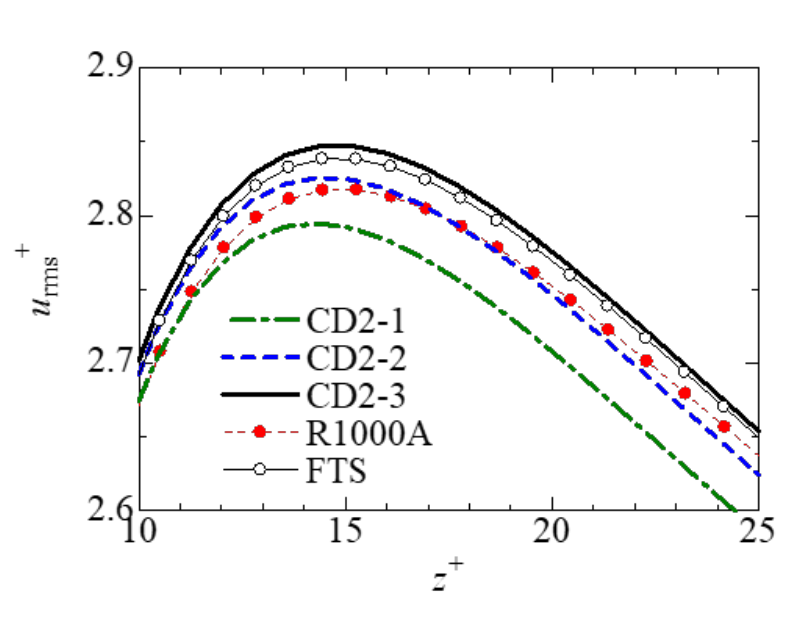}
\caption[ ]{
Comparison of the streamwise turbulence‑intensity distribution $u_{rms}^{+}$ in the vicinity of the near‑wall peak for the CD2 cases, R1000A, and the FTS reference.
}
\label{fig:urms-peak-CD2}
\end{figure}

\begin{figure}
\centering
\includegraphics[width=.7\textwidth]{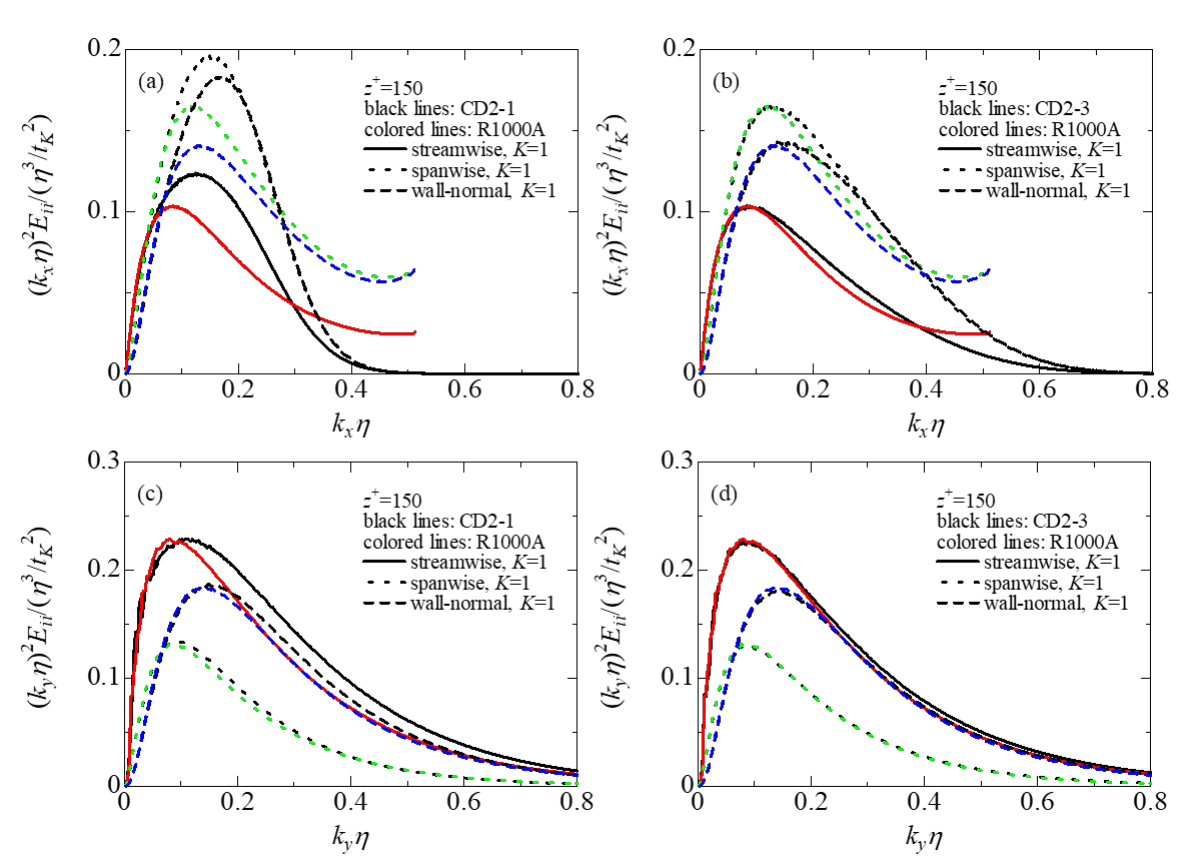}
\caption[ ]{
Comparison of the dissipation spectra for the CD2 cases and the reference spectral solution R1000A. (a) and (b) show the streamwise spectra for CD2‑1 and CD2‑3, respectively, while panels (c) and (d) present the corresponding spanwise spectra.
}
\label{fig:spect-CD2}
\end{figure}

\begin{figure}
\centering
\includegraphics[width=.7\textwidth]{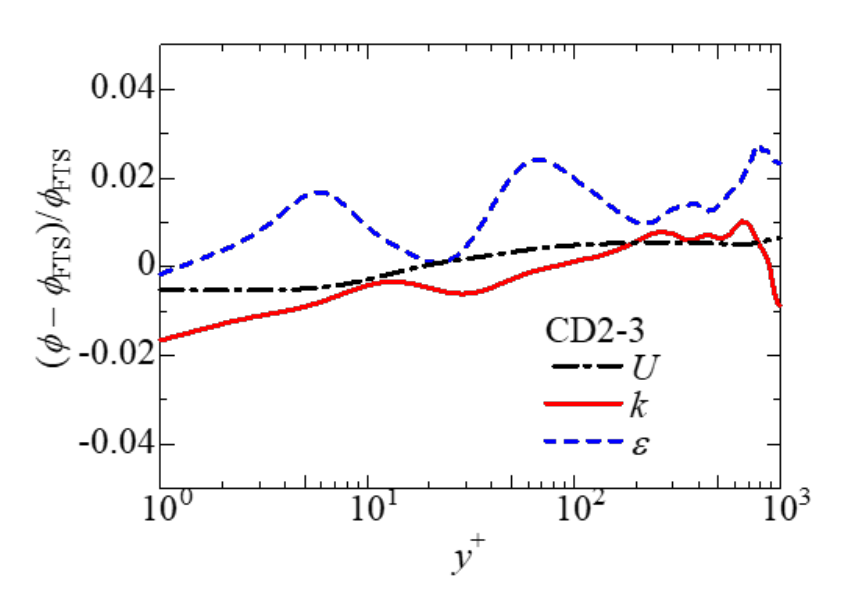}
\caption[ ]{
Relative differences between CD2-3 and the FTS reference for the mean velocity $U$, turbulent kinetic energy $k$, and dissipation rate $\varepsilon$, plotted as $(\phi - \phi_{FTS})/\phi_{FTS})$ against the wall‑normal coordinate $z^{+}$.
}
\label{fig:error-CD2-3}
\end{figure}
The present DNS code employs a spectral method in the wall‑parallel directions and a second‑order central difference scheme in the wall‑normal direction, with the latter resolving the Kolmogorov length scale. One of the advantages of this formulation is that it allows the influence of the wall‑parallel resolution in second‑order central‑difference‑based DNS to be evaluated independently, without contamination from wall‑normal discretization errors. In this chapter, we quantify the wall‑parallel resolution required to achieve an accuracy equivalent to the first‑approximation resolution (R1000A), assuming the use of a fully second‑order central difference scheme in all spatial directions—a configuration that is among the easiest to implement for massively parallel DNS.

\subsection{Overview of the Second‑Order Central Difference Scheme}
In this chapter, we employ the fully second‑order central‑difference DNS code developed by Yamamoto et al.\cite{yamamoto2001turbulence}. The code uses a staggered‑grid arrangement and discretizes the gradient‑type convective terms with the second‑order central‑difference Morinishi scheme\cite{morinishi1998fully}. The time‑integration and pressure‑solution procedures are identical to those used in the FTS code. Consequently, any differences in the numerical accuracy between the two solvers arise solely from the wall‑parallel spatial discretization.

\subsection{Method for Estimating the Equivalent Resolution}
In the second‑order central difference (CD2) scheme, the discretization of spatial derivatives introduces a dispersive truncation error that reduces the effective wavenumber, particularly at high wavenumbers. As a result, a finer grid is required for CD2 to achieve an accuracy comparable to that of the spectral discretization used in the reference DNS. 
In the present study, we begin with a resolution comparable to Pirozzoli et al.\cite{pirozzoli2021natural} $(\Delta x^{+} \approx 10, \Delta y^{+} \approx 5)$ and progressively refine the wall‑parallel grid spacing. By comparing the dissipation spectra and fundamental turbulence statistics with those of the first‑approximation resolution (R1000A), the wall‑parallel resolution required for CD2 to reproduce the same level of accuracy is determined through numerical experiments. The computational conditions used for this assessment are summarized in Table \ref{tab:CD2}.

\subsection{Validation of Second‑Order Central Difference Resolution}
Figure \ref{fig:CD2} compares (a) the mean velocity gradient, (b) the streamwise turbulence intensity, (c) the dissipation rate of the streamwise turbulence intensity, and (d) the dissipation rate of the wall‑normal turbulence intensity for the CD2 cases and the reference solution R1000A. In CD2‑1 and CD2‑2, noticeable deviations from R1000A appear for wall‑normal locations $z^{+} > 40$. This trend is similar to that observed in the under‑resolved spectral case R1000D $(\Delta x^{x} \approx 28.5, \Delta y \approx 15.2)$ as shown in Figs.\ref{fig:R1000ACD} and \ref{fig:e-R1000ACD}. However, unlike R1000D, which exhibits discrepancies even in the near‑wall region $(z^{+} < 15)$, both CD2‑1 and CD2‑2 agree well with R1000A close to the wall, indicating that the impact of insufficient resolution depends strongly on the wall‑normal location.
As shown in Figure \ref{fig:CD2}(b), the streamwise turbulence intensity in CD2‑1 and CD2‑2 exhibits noticeable discrepancies from R1000A, particularly around the near‑wall peak at $z^{+} \approx 7$ and in the range $20 < z^{+} < 100$. Although not shown here, the wall‑normal and spanwise turbulence intensities display comparatively smaller deviations, whereas the total dissipation shows clear discrepancies in the range $20 < z^{+} < 100$ (figures omitted). These results indicate that, within the present CD2 configurations, insufficient wall‑parallel resolution manifests primarily through a distortion of the streamwise‑component dynamics and the associated dissipation distribution in the buffer and lower logarithmic layers, rather than as a uniform degradation of all turbulence statistics.
In contrast, CD2‑3 shows nearly perfect agreement with R1000A for both the mean quantities and the turbulence statistics. This indicates that, in order for the second‑order central difference scheme to achieve an accuracy equivalent to the first‑approximation resolution (R1000A), a wall‑parallel resolution of approximately $\Delta x^{+} \approx 5.0$, $\Delta y^{+} \approx 4.5$ is required. This corresponds to an increase of roughly six times in the number of grid points compared with R1000A.
Figure \ref{fig:urms-peak-CD2} compares the streamwise turbulence‑intensity distribution in the vicinity of the near‑wall peak. Both CD2‑1 and CD2‑2 underpredict the peak value, whereas CD2‑3 slightly overpredicts it. Despite this small overprediction, CD2‑3 is closer to the FTS result than R1000A, indicating that the reduction of turbulent fluctuations below the cutoff wavenumber is almost negligible in CD2‑3. The friction Reynolds number of CD2‑3 is slightly lower ($Re_{\tau} = 994$) than those of FTS and R1000A ($Re_{\tau}$=999 and 1000, respectively). Although CD2‑3 reproduces the near‑wall peak of $u_{rms}$ more accurately than R1000A, this reduction in $Re_{\tau}$ may contribute to the remaining discrepancies observed away from the peak. This indicates that, while CD2‑3 achieves sufficient resolution to capture the peak turbulence intensity, its overall accuracy may still fall slightly short of that of R1000A due to the lower friction Reynolds number.
Figure \ref{fig:spect-CD2} compares the streamwise and spanwise dissipation spectra at $z^{+} \approx 150$, a wall‑normal location where resolution effects are known to be most pronounced (see Figures \ref{fig:resolution-A} and \ref{fig:resolution-DIS}). In CD2‑1, a clear overprediction is observed from the dissipation peak down to $k_{x}\eta$ or $k_{y}\eta < 0.4$, which directly corresponds to the deviations in the dissipation rates shown in Fig.\ref{fig:CD2}. In contrast, CD2‑3 exhibits only a slight overprediction in the streamwise direction (Figure \ref{fig:spect-CD2}(b)), while the spanwise spectrum (Figure \ref{fig:spect-CD2}(c)) matches R1000A almost perfectly. These results demonstrate that accurate reproduction of the dissipation rates requires agreement at the level of the dissipation spectra, particularly in the high‑wavenumber range where resolution effects are most sensitive.
Following the above results, CD2‑3, which employs approximately six times more grid points than R1000A, achieves comparable accuracy to R1000A not only in the mean quantities and turbulence statistics but also in the dissipation spectra. Moreover, CD2‑3 reproduces the near‑wall peak of $u_{rms}$ even more closely than R1000A, showing excellent agreement with the FTS reference. However, as discussed earlier, CD2‑3 exhibits a slight underprediction of the wall shear stress, resulting in a lower friction Reynolds number ($Re_{\tau} = 994$). 
The relative‑error representation with respect to the FTS reference, shown in Fig. \ref{fig:error-CD2-3}, indicates that CD2‑3 exhibits a noticeable deviation in the mean velocity gradient even in the near‑wall region. This deviation is consistent with the reduced friction Reynolds number ($Re_{\tau} = 994$) compared with the FTS value ($Re_{\tau} = 994$). Although CD2‑3 reproduces the near‑wall peak of $u_{rms}$ more closely than R1000A, the relative‑error plots for the turbulent kinetic energy and dissipation rate clearly show larger discrepancies than those of R1000A, as seen in Fig. \ref{fig:error-R1000A}. These results demonstrate that CD2‑3 does not reach the overall accuracy of R1000A, despite its finer wall‑parallel resolution.
A notable finding of the present analysis is that the resolution requirement of the CD2 scheme is substantially more stringent in the streamwise direction than in the spanwise direction. Conventional CD2‑based DNS often employ wall‑parallel resolutions on the order of $\Delta x^{+} \approx 8$, $\Delta y^{+} \approx 4$(see e.g., \cite{diez2025pencil}) , which have generally been regarded as sufficient. However, the present results show that such resolutions are inadequate for accurately capturing the high‑wavenumber streamwise fluctuations when compared with the spectral reference.
This increased sensitivity is related to the wavenumber dependence of the phase speed of turbulent fluctuations. In the streamwise direction, the phase speed of small‑scale motions increases toward higher wavenumbers, whereas in the spanwise direction it decreases with increasing wavenumber (see, e.g., \cite{del2009estimation, bernardini2013turbulent}). In contrast, the second‑order central difference scheme introduces numerical dispersion that reduces the phase speed at high wavenumbers in all directions. As a result, the numerical dispersion error acts in the same direction as the physical trend for high‑$k_y$ modes, but in the opposite direction for high‑$k_x$ modes, leading to a much stronger mismatch between the numerical and physical phase speeds in the streamwise direction. This mismatch enhances the attenuation of high‑$k_x$ components and explains why significantly finer streamwise resolution is required for CD2 to reproduce the spectral reference.

\section{\label{sec:level8}Discussion and Conclusion}
The present study conducted a Full Turbulence Simulation (FTS) of turbulent channel flow at $Re_{\tau} \approx 1000$, resolving the Kolmogorov wavenumber in all spatial directions. The simulation was validated through dissipation spectra and detailed comparisons with established DNS data, confirming that the present computation satisfies the criteria for FTS. Based on this fully resolved reference, the spatial resolution required for high‑Reynolds‑number DNS was quantitatively examined.
A distinctive feature of the present FTS at $Re_{\tau} \approx 1000$ is that it resolves a wall‑normal region that does not exist in the lower‑Reynolds‑number DNS commonly used for turbulence‑model validation, such as the well‑established cases at $Re_{\tau} \approx 180$\cite{kim1987turbulence} and 395\cite{moser1999direct}. According to Eq. (\ref{eq:range}), the intermediate layer occupies only a negligible range at these Reynolds numbers, whereas at $Re_{\tau} \approx 1000$ it attains a physically meaningful width of $\mathcal{O}(100)$ viscous units. The present FTS fully resolves this region, capturing the onset of the intermediate‑layer dynamics that become increasingly important at higher Reynolds numbers. This makes $Re_{\tau} \approx 1000$ the lowest Reynolds number at which reliable resolution requirements for high‑Re DNS can be established. The main findings obtained from the present FTS and the subsequent resolution analysis are summarized as follows.
\begin{itemize}
\item On the use of a second‑order central‑difference scheme in the wall‑normal direction
\end{itemize}
As demonstrated in Chapter\ref{sec:level2}, when the computational domain and the wall‑parallel resolution are matched, the use of a second‑order central‑difference scheme in the wall‑normal direction introduces no adverse effects on the accuracy of DNS. The comparisons showed that the mean velocity, turbulence intensities, and dissipation characteristics are reproduced with high fidelity, indicating that the second‑order scheme is fully adequate for resolving the wall‑normal dynamics at the present grid spacing. These results clarify that the concerns raised by Nagib et al.\cite{nagib2024utilizing} regarding the use of second‑order central differences arise from under‑resolved configurations and do not apply when the wall‑normal grid spacing is sufficiently fine.
\begin{itemize}
\item On the use of moderately coarse wall‑parallel grids in spectral DNS
\end{itemize}
Regarding the use of moderately coarse wall‑parallel grids in spectral discretization, the present results show that such resolutions remain fully adequate for capturing the essential turbulence dynamics. As demonstrated in Chapter \ref{sec:level5}, grid spacings of $\Delta x^{+} \approx 17.8$ and $\Delta y^{+} \approx 7.4$(R1000A) resolve more than 99\% of the turbulent kinetic energy and accurately reproduce the dominant contributions to the dissipation terms.
The only noticeable limitation at this resolution is a slight underprediction of the near‑wall peak of $u_{rms}$, which arises from the small amount of turbulent energy residing below the cutoff wavenumber. This discrepancy is confined to the peak region and does not affect the near‑wall cycle, the mean velocity gradient, or the major turbulence statistics. These findings clarify that the use of moderately coarse spectral grids does not compromise the fidelity of DNS at $Re_{\tau} \approx 1000$, and that the resolution sensitivity observed in the intermediate layer reflects physical scale separation rather than any limitation of the spectral method itself.
\begin{itemize}
\item On the DNS required to reproduce the near‑wall $u_{rms}$ peak
\end{itemize}
The present results also indicate that the near‑wall peak of $u_{rms}$ is influenced not only by spatial resolution but also by the size of the computational domain. As shown in Chapter \ref{sec:level5}, small differences in the domain size can lead to slight variations in the peak intensity, as evidenced by the fact that the LM dataset $(L_{x}=8\pi h, L_{y}=3\pi h)$ \cite{lee2015direct} yields a marginally higher peak than the FTS$ (L_{x}=16h, L_{y}=6.4h)$ despite its coarser resolution. In contrast, Chapter \ref{sec:level6} shows that when the computational domain is matched $ (L_{x}=16h, L_{y}=6.4h)$, the LM‑level resolution ($\Delta x^{+} \approx 10$ and $\Delta y^{+} \approx 5$) produces a slight underprediction relative to the fully resolved FTS ($\Delta x^{+} \approx 4$ and $\Delta y^{+} \approx 4$), indicating that the LM resolution is not necessarily robust for this quantity. These observations imply that establishing a reliable DNS requirement for reproducing the $u_{rms}$ peak demands both a computational domain larger than that of LM and a wall‑parallel resolution comparable to the FTS conditions.
\begin{itemize}
\item On the use of second‑order central differences in the wall‑parallel directions
\end{itemize}
Chapter \ref{sec:level7} shows that applying a second‑order central‑difference scheme in the wall‑parallel directions requires reproducing the dissipation‑spectrum distribution with fidelity comparable to the spectral method. Achieving this level of accuracy demands wall‑parallel grid spacings of approximately $\Delta x^{+} \approx 5.0$ and $\Delta y^{+} \approx 4.5$, corresponding to the CD2‑3 case. However, even with this high‑resolution grid, the second‑order scheme exhibits a clear degradation in accuracy compared with the moderately coarse spectral‑method case R1000A ($\Delta x^{+} \approx 17.8$ and $\Delta y^{+} \approx 7.4$). This result indicates that, in the wall‑parallel directions, the second‑order scheme cannot robustly match the accuracy of the spectral method, and that substantially finer grids are required merely to approach the fidelity obtained with a much coarser spectral discretization. It should be noted that the requirements for the wall‑parallel directions differ fundamentally from those for the wall‑normal direction, where the second‑order scheme was shown to be fully adequate at the present resolution, which sufficiently resolves the Kolmogorov length scale.

These findings also suggest that, for reproducing the near‑wall peak of $u_{rms}$, some degree of compromise is unavoidable in DNS at $Re_{\tau} = \mathcal{O}(10^{4})$ and beyond. Although the R1000A resolution underpredicts the peak intensity, it uses only one‑eighth of the grid points required for the FTS and still reproduces the essential turbulence statistics within 1\% accuracy. In contrast, applying a second‑order central‑difference scheme in the wall‑parallel directions requires grid spacings approximately six times finer than R1000A, yet even such high‑resolution configurations fail to reach the accuracy of the moderately coarse spectral case. This clearly indicates that the wall‑parallel discretization becomes a major bottleneck under the memory constraints of high‑Reynolds‑number DNS. The resolution criteria derived here therefore offer practical guidance for designing high‑Reynolds‑number simulations, including those at $Re_{\tau} = \mathcal{O}(10^{4})$, where balancing computational cost and physical fidelity remains a central challenge.

\begin{acknowledgments}
his work was partly supported by JSPS KAKENHI (C) Grant No. JP23K03656. Computational resources were provided by the Fugaku supercomputer at the RIKEN Center for Computational Science through the HPCI System Research Project (Project IDs: hp200189, hp210125, hp220144, hp230138, hp240171, hp250159). Part of the results were obtained using supercomputing resources at the Cyber science Center, Tohoku University, and through the Collaborative Research Project for Large‑Scale Computation at the Academic Center for Computing and Media Studies, Kyoto University. This research also used the Fujitsu PRIMEHPC FX1000 system (Wisteria/BDEC‑01) under the “Large‑scale HPC Challenge” Project at the Information Technology Center, The University of Tokyo.

\end{acknowledgments}

\appendix
\section{Convergence of the Total Shear Stress}
\label{app:convergence}
\begin{figure}
\centering
\includegraphics[width=.7\textwidth]{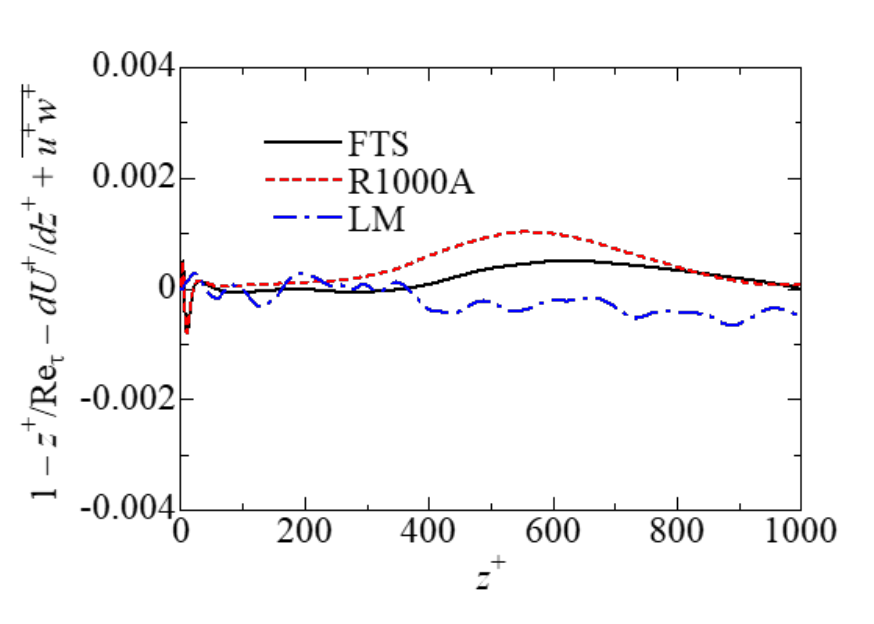}
\caption[ ]{Deviation of the total shear stress from its ideal linear distribution
}
\label{fig:convergence}
\end{figure}

Figure \ref{fig:convergence} shows the deviation of the total shear stress from its ideal linear distribution, $1 - z^{+}/Re_{\tau} - dU^{+}/dz^{+} +\overline{ u^{+}w^{+} }$. The maximum deviation remains below $10^{-3}$ across the entire channel, demonstrating that the mean momentum balance is accurately satisfied and that the statistical convergence of the present simulations is sufficient.

\appendix
\bibliographystyle{apsrev4-2}
\bibliography{FTS.bbl}
\end{document}